\documentclass[12pt,4paper,oneside]{article}

\usepackage[T1]{fontenc}   
\usepackage[brazilian,english]{babel} 
\usepackage[utf8]{inputenc}
\usepackage{array}
\usepackage{natbib}     
\usepackage{amsmath}     
\usepackage{amssymb}     
\usepackage{graphicx}    
\usepackage{fullpage}    
\usepackage{setspace}	 
\usepackage{float}      
\usepackage{booktabs}   
\usepackage{rotating}
\usepackage{caption}
\usepackage{adjustbox}
\usepackage{booktabs,caption}
\usepackage[flushleft]{threeparttable}
\usepackage{titling} 
\newcommand\fnote[1]{\captionsetup{font=footnotesize}\caption*{#1}} 
\global\long\def\sym#1{\ifmmode^{#1}\else$^{#1}$\fi}%

\usepackage[hidelinks]{hyperref}
\usepackage[usenames,dvipsnames]{xcolor} 
\hypersetup{
  colorlinks   = true, 
  urlcolor     = blue, 
  linkcolor    = blue, 
  citecolor   = blue 
}

\usepackage{newtxtext,newtxmath}
\usepackage[top=25.4mm, bottom=25.4mm, left=25.4mm, right=25.4mm]{geometry} 

\begin{document}

\title{A Lifecycle Estimator of \\ Intergenerational Income Mobility\thanks{We thank Adrian Adermon, Nathan Deutscher, Paul Hufe, Alejandro Puerta, Nazarii Salish, Felix Wellschmied, and conference and seminar participants at SOFI, the ``Inequality, Education and Mobility Workshop'', Uppsala University, the ifo ``Opportunities, Mobility and Well-being Workshop'', IAE-CSIC, the 2021 EALE, PUC-Rio, Insper, USP and PUC-Chile for comments. We thank Juan C. Palomino for generously sharing data preparation codes for the PSID. Support from the Ministerio de Ciencia e Innovación (Spain, RYC2019-027614-I and CEX2021-001181-M) and Comunidad de Madrid (AM-EPUC3M11) is gratefully acknowledged. This manuscript has been accepted at the Review of Economics and Statistics, https://doi.org/10.1162/rest\_a\_01585.}}

\author{Ursula Mello\thanks{Insper Institute of Education and Research, Institute for Economic Analysis and IZA.} \and Martin Nybom\thanks{IFAU and UCLS.}
\and Jan Stuhler\thanks{Department of Economics, Universidad Carlos III de Madrid.}}

\vspace{1.5\baselineskip}\vspace{-\parskip}

\date{December 2025}

\maketitle
\addtocounter{page}{-1}
\thispagestyle{empty}
\doublespacing

\begin{abstract}\onehalfspacing
Lacking lifetime income data, most intergenerational mobility estimates are subject to lifecycle bias. Using long income series from Sweden and the US, we illustrate that standard correction methods struggle to account for one important property of income processes: children from affluent families experience faster income growth, even conditional on their own characteristics. We propose a lifecycle estimator that captures this pattern and performs well across different settings. We apply the estimator to study mobility trends, including for recent cohorts that could not be considered in prior work. Despite rising income inequality, intergenerational mobility remained largely stable in both countries.
\vspace{0.5\baselineskip}\vspace{-\parskip}


\end{abstract}

\newpage
\doublespacing

\section{Introduction} \label{intro}

A key statistic to characterize inequality is the intergenerational elasticity of income (the IGE), an inverse measure of intergenerational mobility. But while the IGE is ideally based on \textit{lifetime} incomes for two generations, most studies have to rely on short snapshots at specific ages. The main challenge, therefore, is to account for the measurement errors introduced by the use of such snapshots. While the literature has made progress on this front, concerns persist about the robustness of available estimates and the reliability of comparisons of IGEs across place or time (\citealt{mogstad2021}). For example, recent US estimates of the IGE range from 0.35 to 0.65, despite building on similar methodological insights (e.g., \citealt{Chetty2014}, \citealt{Mazumder2016}). 

We propose a new \textit{lifecycle} estimator of the IGE in incomplete income data that is less sensitive to the age at which child income is measured. Our estimator models income profiles as a function of age and education, but also allows income growth to vary with parental background conditional on own characteristics. We use long income series from Swedish registers and the Panel Study of Income Dynamics (PSID) to illustrate that these family effects on income growth are sizable, and to verify that the estimator performs well across different data settings. We then apply the estimator to study trends in intergenerational mobility for cohorts born between the 1950s and 1980s, in both the US and Sweden.

We start by analyzing the key components of the income process that affect intergenerational estimators: (i) income growth explained by an individual's own characteristics, (ii) transitory noise, and (iii) income growth unexplained by own characteristics. Crucially, this unexplained income growth correlates within families: within education or occupation groups, children from more affluent families tend to have lower initial incomes but steeper income growth. For example, college-educated sons with fathers from the top quartile of the Swedish income distribution earn less in their mid-20s, yet have around 40 percent higher incomes by age 40, compared to college-educated sons from bottom-quartile families. These findings matter for the estimation of income mobility but also for the broader debate on the properties of income processes, supporting the notion that income grows at an individual-specific and deterministic rate (\citealt{Guvenen2009}).

We then analyze whether existing methods account for these properties of the income process. Two strategies can be distinguished. One option is to formalize the relation between (observed) annual and (unobserved) lifetime income in an errors-in-variables model, as in \cite{Haider2006}. Alternatively, one may estimate the shape of income profiles over the lifecycle, using partially observed profiles and observable characteristics, as in \cite{Hertz2007}. Following the first approach, most applications seek to reduce lifecycle bias by measuring incomes around midlife. Although this rule of thumb is useful, the ``ideal'' age to minimize bias is typically unknown, and the selected age ranges vary considerably across studies (see Table A.1). Moreover, midlife incomes are not observed for recent birth cohorts. 

Our proposed estimator exploits the available income information more fully. We first estimate income profiles based on standard observables, such as age and education (as in \citealt{Vogel2007} and \citealt{Hertz2007}), to predict individual income at unobserved ages. However, we also account for the fact that children from affluent families experience steeper growth \textit{conditional} on those observables, and show that this reduces the sensitivity of mobility estimates to the age at which income is observed. Specifically, by controlling for the interaction between parental income and child age, we reduce the correlation between the prediction error in child income and parental income – a key condition for unbiased estimation of the IGE. Alternatively, we allow the child's income growth to depend on their income level, to capture the ``fanning out'' of income profiles over age (\citealt{Creedy1988}). We then predict complete lifecycle income profiles and estimate the IGE in lifetime income. 

In other words, before estimating the association of parent to child income (the intergenerational regression), we first study its relation to the child's income \textit{growth}. We show that the proposed lifecycle estimator performs well in both Swedish and US data, closely tracking a benchmark estimate based on lifetime incomes. The estimator is largely robust to the exact age range or the number of income observations available per individual and, therefore, promises to be applicable in a wide range of settings. In contrast to current practice, it makes use of all available income information. The estimator is particularly useful for analyzing mobility for recent cohorts, whose incomes are observed only at younger ages. Our main analysis thus relies on a parametric first-step model, including predictors that are known sources of lifecycle bias and commonly available in standard data sets. However, since this first step is essentially a prediction problem, alternative data-driven approaches might perform as well or better. We show that in our data simple plug-in estimators tend to perform worse than our parametric specifications, but future research could explore the potential of more tailored implementations. 

Finally, we apply the estimator to study mobility trends in Sweden and the US. Our objective is three-fold. First, we examine whether prior estimates may be systematically biased. Second, we estimate mobility trends for younger, more recent birth cohorts -- which are especially relevant from a policy perspective -- exploiting that our method works well even if incomes are observed only at an early age. Third, this application moves beyond the `'idealized data setting'' used in the preceding analyses, making it a more practical reference for other researchers interested in applying the estimator. For Sweden, accounting for lifecycle effects leads to different conclusions on how mobility has developed over time. Basic fixed-age estimates suggest that mobility declined sharply between the 1950s and 1970s cohorts. Accounting for lifecycle effects, however, yields much more stable mobility estimates over these cohorts, and a slight increase for those born in the 1980s. Thus, Sweden's comparatively high level of income mobility has remained a persistent feature in the second half of the 20th century.  

For the US, failing to account for lifecycle dynamics leads to an overestimation of income mobility for cohorts born in the 1980s, who are only observed at young ages. A naive estimator would thus falsely suggest a sharp increase in mobility. Our lifecycle estimator produces larger and more stable IGE estimates, suggesting that mobility remained remarkably constant over recent decades in the US. 
Our findings therefore contrast with the hypothesis that mobility must have plunged for children born in the 1980s (\citealt{putnam2012growing}), and the observation that socioeconomic gaps in parents' monetary and time investments did increase (\citealt{ramey10}, \citealt{Corak2013}). 
An interesting question for future work is why mobility in  \textit{outcomes} has remained stable despite widening gaps in \textit{inputs}.

Our paper adds to an extensive literature on IGE levels and trends across countries (\citealt{Solon1999}; \citealt{Black2011}; \citealt{Jenkins2015})
and on measurement error in intergenerational estimates. Early studies focused on classical errors from incomplete income data for \textit{parents} (\citealt{Atkinson1980}, \citealt{Solon1999}), recognizing that lifecycle variation should be accounted for but assuming that basic age controls in IGE regressions would suffice. Subsequent studies addressed non-classical measurement error, shifting the attention to lifecycle bias from incomplete data for the \textit{child} generation. First discussed in \cite{Jenkins1987}, the problem gained wider attention following \cite{Haider2006} and applications by \cite{Grawe2006},
\cite{Bohlmark2006}, and \cite{Nilsen2012}, among others.\footnote{In parallel, \cite{Creedy1988}, \cite{Vogel2007} and \cite{Hertz2007} suggested alternative approaches based on modelling how the income profile varies with observable characteristics, and \cite{Chau2012} and \cite{Jantti2012} consider lifecycle models with heterogeneous intercepts and slopes.} In light of these issues, recent work often favors rank-based measures that are less sensitive to measurement problems (e.g., \citealt{Chetty2014}).\footnote{See also \cite{Nybom2017} for a comparison of different mobility measures, and \cite{Kitagawa2019} for a correction method for measurement error in ranks.} As these measures capture positional dependence but abstract from variation in income inequality, the IGE remains a key measure in the analysis of income mobility.

Our method is particularly suitable for studying mobility variations across countries or over time (\citealt{Blanden2013}, \citealt{Corak2013}, \citealt{Chetty2014}). Comparisons of the IGE across different contexts are difficult: if the shape of income profiles differs, measuring incomes at a specific age will introduce biases of varying magnitudes or even signs. This problem also arises when studying mobility \textit{trends}. For Sweden the evidence is scarce, and we present credible estimates for cohorts born through the 1980s. The US literature is more extensive. Earlier work found no evidence of shifts in mobility for cohorts born up until around 1980 (e.g. \citealt{Hertz2007}, \citealt{Lee2009}, \citealt{Chetty2014a}) -- a surprising finding given the concurrent increase in income inequality, the theoretical and empirical link between inequality and mobility (\citealt{Blanden2013}, \citealt{Corak2013}, \citealt{Chetty2014}), and increasing gaps in parental inputs (\citealt{blandendoepkestuhlerHB2022}). However, \cite{Davis2019} found that mobility declined sharply for earlier cohorts not well covered by the PSID, while \cite{Justman2021} and \cite{jacome2021mobility} found declining mobility for cohorts also considered by \cite{Hertz2007} and \cite{Lee2009}. We argue that the consideration of family effects on income growth may help to produce more comparable estimates, including for recent cohorts that have not yet been studied.

Our arguments also relate to the discourse on income processes (e.g., \citealt{Meghir2011}). This link is interesting in both directions. On the one hand, a better understanding of income processes may help researchers to assess the usefulness of different intergenerational estimators. On the other hand, an intergenerational perspective may inform the debate on the role of unobserved heterogeneity in income processes. A controversial question is whether (residual) income grows at an individual-specific and deterministic rate or follows a random walk. Distinguishing between these models is difficult, and conventional tests of the covariance structure of income growth may not be very informative \citep{Guvenen2009}.  We argue that intergenerational data provide evidence in favor of the model with heterogeneous growth rates – within education or occupational groups, children from affluent families experience substantially faster income growth than those from low-income families, particularly early in their career (see also \citealt{Michele2018}, \citealt{halvorsen2021earnings}, and \citealt{lochner2022earnings}). 

The paper's sections are divided as follows. In Section \ref{sec:data}, we describe the Swedish and US data and our sampling. Section \ref{sec:incprocess} provides a discussion of the properties of the income process and evidence on its key components. In Section \ref{sec:correctionmethods}, we analyze existing correction methods for IGE estimates in light of the income process properties. Section \ref{sec:newcorrection} presents and tests our new estimator, which we use in Section \ref{sec:correctiontrends} to study mobility trends in Sweden and the US. Section \ref{sec:conclusion} concludes.

\section{Data} \label{sec:data}

We use data from Sweden and the US. For Sweden, we use various administrative registers that contain the universe of Swedish citizens aged 16-64 at any point between the years 1960-2018 (born 1896-2002) and gross labor earnings from tax records (reported by employers) for the period 1968-2018.\footnote{See \cite{SCB_2019c, SCB_2019a, SCB_2019f, SCB_2019e, SCB_2019b, SCB_2019d, NA_2005}. We observe labor earnings (including from self-employment) for all residents in 1968, 1970, 1971, 1973, 1975, 1976, 1979, 1980, 1982 and annually for 1985-2018. We impute data for the gap years that occurred after 1968 with neighboring observations, bottom code annual incomes to 10,000 SEK (around 912 USD) and adjust incomes for inflation using the CPI.} Using multigenerational registers, we link children born 1932 or later (who have been residents of Sweden at some point since 1961) to their biological parents. As we observe individual rather than household income, we focus on father-son pairs to abstract from female labor market participation, which also improves comparability with the previous literature. Other administrative registers provide information on education, occupation and further individual characteristics.\footnote{The Education Register contains data on highest educational attainment and field of education for practically the entire population alive in 1970 or later. Occupational information comes from different registers and is available bidecennially from 1960 to 1990 and, for a large subsample, annually from 1996 and onwards.} 

We construct two different samples: a \emph{benchmark sample} for studying the performance of different estimators and a \emph{trends sample} for studying trends in income mobility. The benchmark sample is chosen such that it contains nearly complete income trajectories, allowing us to compare estimates from partial data to a ``true'' benchmark estimate based on lifetime incomes. Specifically, we consider cohorts born 1952-1960, with incomes observed between ages 25 and 58. To measure fathers' income over a long period (age 41-58), we drop fathers born before 1927. The implied drop in representativeness is not a concern for our purposes, provided the sample remains sufficiently diverse to study lifecycle patterns in income. 

Our trends sample covers cohorts born 1950-1989, which we analyze by decade of birth. To ensure that the parental income measure is comparable across cohorts, we construct it in two steps. First, we randomly select up to five annual income observations for each father (between age 40 and 55), such that the number of observations per father is similar across cohorts. Second, we use these annual observations in an extended Mincer-type equation to predict income at age 50 for each father. Specifically, we regress  annual incomes on individual fixed effects and a cohort-by-education specific age polynomial (of the father). We therefore balance both the number of income observations and the age at measurement across cohorts.

For the US, we use data from the Panel Study of Income Dynamics (\citealt{psid2019}), which began in 1968 with a nationally representative sample of over 18,000 individuals living in 5,000 families. The survey is useful for intergenerational research since it follows children from the original sample as they form their own households, and contains data on employment, income, and education.  
Apart from a few exceptions, we follow the sampling and variable definitions in \cite{Lee2009}. As such, we use only the PSID core sample (the Survey Research Center component). We focus on family rather than individual income and pool sons and daughters to maximize our sample. 

Our benchmark sample covers children born 1952-1960, which similarly to the Swedish case enables us to observe almost complete income histories. To measure parental income, we average log annual family income in the childhood home over the three years when the child was 15-17 years old, similar to the measures in \cite{Lee2009} and \cite{Chetty2014}. We measure the children's adult income by the (log) annual family income in the household in which they were the household head or head's spouse and exclude outlier observations (using the same thresholds as \citealt{Lee2009}). As for Sweden, we also construct a trends sample covering US cohorts born 1950-1989, using the same sampling and variable definitions as for the benchmark sample.

To improve comparability with previous studies, we use similar sampling and variable definitions as in previous work. As such, those definitions differ between the two countries. First, we use family income and consider both sons and daughters for the US, while for Sweden we consider individual labor earnings and father-son pairs. Second, we measure parental income at a given age of the child for the US, but at a given age of the parent for Sweden. Third, the parental income measure is based on up to 18 years of income for Sweden but a three-year average for the US, such that the US estimates are more strongly attenuated by measurement error (see \citealt{Mazumder2005}). For these reasons, we cannot directly compare mobility levels between countries. However, our results are comparable to prior work for each country, as well as across cohorts within countries.

\begin{table}[h!]
  \centering
  \caption{Descriptive Statistics}
    \begin{adjustbox}{width=\textwidth}
    \begin{tabular}{lccccc}
    \toprule
    \toprule
          & Benchmark & 1950-59 & 1960-69 & 1970-79 & 1980-89 \\
    \midrule
          & \multicolumn{4}{c}{\textit{Panel A: Swedish Register Data }} &  \\
    \midrule
    Father-son pairs & 201,066 & 525,813 & 572,811 & 532,835 & 530,312 \\
    Father's Age at Birth of Son & 25.7  & 31.6  & 30.1  & 29.6  & 31.1 \\
    Log Lifetime Income of Sons & 12.4  & 12.4  & 12.5  & 12.5  & 12.3 \\
    \% Zero Income Obs of Sons & 8.6   & 8.2   & 8.7   & 8.7   & 11.2 \\
    Mean Age Son at Earnings Obs & 40.9  & 42.4  & 39.1  & 34.5  & 29.6 \\
    \% Sons with College Degree & 14.1  & 16.2  & 17.0  & 27.1  & 26.2 \\
    \% Sons in Managerial Position & 17.8  & 17.9  & 17.9  & 13.1  & 5.1 \\
    Log Lifetime Income of Fathers & 12.3  & 12.3  & 12.3  & 12.3  & 12.4 \\
    Percent of Zero Income Obs of Fathers & 6.0   & 5.4   & 6.2   & 6.9   & 8.6 \\
    \% Fathers with a College Degree & 7.7   & 7.3   & 10.6  & 14.9  & 16.8 \\
    \% Fathers in Managerial Position & 11.8  & 11.6  & 14.7  & 16.3  & 12.3 \\
     \midrule
          & \multicolumn{4}{c}{\textit{Panel B: PSID}} &  \\
    \midrule
    Parent-child pairs & 1,286 & 1,283 & 1,153 & 1,212 & 1,419 \\
    Share Female & 51.2  & 50.9  & 51.3  & 48.9  & 53.4 \\
    Log Lifetime Income Child & 9.2   & 9.2   & 9.2   & 9.2   & 8.9 \\
    Mean Age Child at Earnings Obs & 36.1  & 36.2  & 34.2  & 31.6  & 27.7 \\
    Log Income Parent when Child was 15-17 & 9.3   & 9.3   & 9.4   & 9.3   & 9.4 \\
    Mean Age Parent when Child was 15-17 & 45.2  & 45.5  & 44.7  & 42.5  & 44.2 \\
    \% Child with Some College & 51.9  & 52.6  & 55.9  & 69.1  & 72.4 \\
    \% Parent with Some College & 26.9  & 26.8  & 35.7  & 50.0  & 56.4 \\
     \bottomrule
      \bottomrule
    \end{tabular}%
  \end{adjustbox}
  \label{tab:tab1}%
  \smallskip 
  \fnote{\textbf{Notes:} The benchmark samples (column 1) contain the cohorts born 1952-1960 and are used for testing the performance of the intergenerational elasticity estimators. The other columns present descriptive statistics from the samples used to study mobility trends, separately by decade of birth of the child.}
\end{table}%

Table \ref{tab:tab1} reports descriptive statistics for each sample. Our benchmark samples contain 201,066 and 1,286 individuals, respectively. The Swedish trends sample covers more than 2 million sons, while the corresponding sample for the US includes approximately 5,000 sons and daughters. Columns 2 to 5 show summary statistics by decade of birth. Unsurprisingly, more recent cohorts are more educated and have more educated fathers, but since their incomes are measured at earlier ages this does not result in higher average income. Our benchmark sample for Sweden is slightly negatively selected in terms of income and education (cf. columns 1 and 2), due to its restriction to younger parents for whom we observe more complete income series. As a consequence, estimates of the IGE will differ between the benchmark sample and the more representative trends sample. 

\section{An Intergenerational Perspective on Income Processes}  \label{sec:incprocess}

We start by illustrating those properties of the income process that are particularly important for intergenerational research. This evidence will then allow us to characterize the advantages and limitations of existing correction methods (Section \ref{sec:correctionmethods}) and to motivate a new lifecycle estimator (Section \ref{sec:newcorrection}) that addresses our key observation – that income growth varies with parental income even conditional on individuals' own characteristics.

A large literature on income processes studies the shape of income profiles over the life cycle.\footnote{Insights from this literature have been used to study the causal effect of parental income (e.g., \citealt{Carneiro2021}), but used only for motivational purposes in descriptive studies (an exception is \citealt{Heidrich2016}).} While many properties are well established, two contrasting viewpoints exist about the idiosyncratic components of income growth. The \textit{restricted income profile} (RIP) model views income as the sum of a mean-reverting component reflecting \textit{transitory shocks} and a random-walk component reflecting \textit{permanent shocks} (\citealt{MaCurdy1982}). In contrast, the \textit{heterogeneous income profile} (HIP) model assumes that incomes grow at an individual-specific and deterministic rate (\citealt{Guvenen2009}). 

The RIP and HIP models are difficult to distinguish in standard data sets, but intergenerational data may provide additional insights. As a reference point, consider the HIP model by \cite{Guvenen2009}, which assumes that log income for individual $i$ with experience $h$ at time $t$ is given by 
\begin{equation}\label{eq:Guvenen_Y}
y^i_{h,t} = g(\theta^0_t,X^i_{h,t}) + f(\alpha^i,\beta^i,X^i_{h,t}) + z^i_{h,t}+\phi_t\varepsilon^i_{h,t}.
\end{equation}
The function \emph{g} captures common income variation
 that is explained by observable characteristics $X^i_{h,t}$.\footnote{\cite{Guvenen2009} considers a cubic polynomial in experience \textit{h}. Yet, more generally, we could think of $X^i_{t}$ as observables that could include education, gender, age, etc. The coefficients $\theta^0_t$ are common to all individuals.} In our analysis, we consider in $X^i_{h,t}$ characteristics that are typically observed by the researcher, such as education or occupation. The second function, \emph{f}, captures the component of life-cycle earnings that is individual or group-specific and that is unexplained by those characteristics. 
By ``unexplained'' we refer to determinants that are \emph{typically} unobserved, such as an individual's ability or parental lifetime income. In the data we use, however, we observe proxies for those characteristics and can therefore test whether they predict lifecycle profiles. 
Finally, the dynamic component of income is modeled as an AR(1) process, $z^i_{h,t}=\rho z^i_{h,t-1}+\pi_t\eta^i_{h,t}$, with $z^i_{0,t}=0$ and with $\pi_t$ capturing possible time-variation in the innovation variance, plus a purely mean-reverting transitory shock, $\varepsilon^i_{h,t}$, scaled by $\phi_t$ to account for possible non-stationarity in that component.\footnote{The innovations  $\varepsilon$ and $\eta$ are assumed to be independent of each other and over time while the vector $(\alpha^i,\beta^i)$ is distributed across individuals with zero mean, variances of $\sigma^2_\alpha$ and $\sigma^2_\beta$, and covariance $\sigma_{\alpha\beta}$. Persistent and transitory shock components are scaled by time-specific coefficients, as they may change over time (\citealt{Moffitt1995}).} 

\begin{figure}[tb]
\caption{\label{fig:fig1} Components of the Income Process in the Swedish Data}
\centering
\includegraphics[width=1\textwidth]{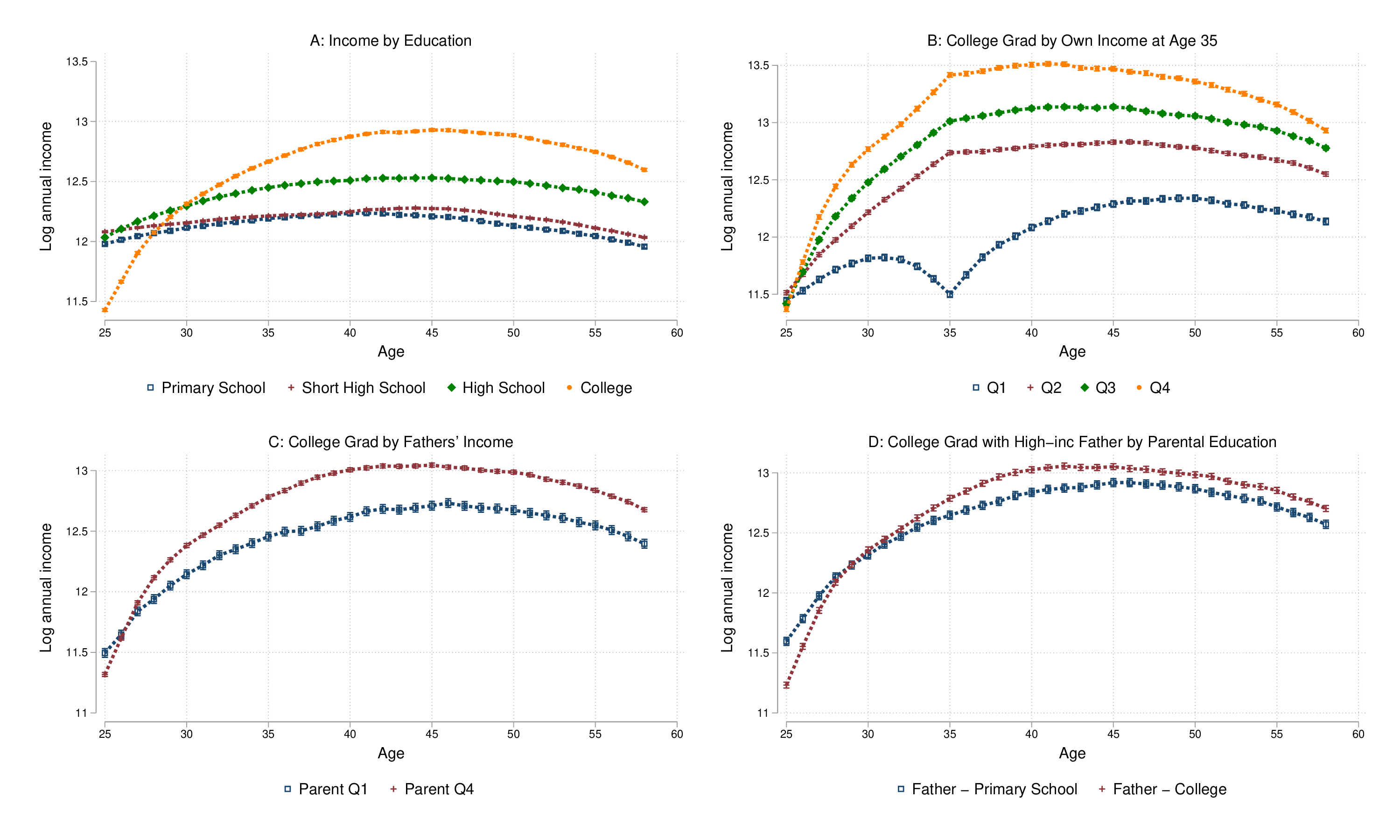}
\fnote{\textbf{Notes:} 
Panel A shows income trajectories by education category. Panel B focuses only on college-educated sons, who are split into four groups according to their annual income at age 35. Category Q1 refers to the bottom and Q4 to the top quartile. In Panel C, college-educated sons are divided into two groups, those in the top and those in the bottom quartile of fathers' lifetime income. In Panel D, college-level sons whose fathers belong to the top half of lifetime income are divided into college-educated fathers and fathers with only primary schooling. We remove time effects to abstract from the business cycle. Confidence intervals (95\%) are plotted around each line.}
\end{figure}

\begin{figure}[tb]
\caption{\label{fig:incomeprocPSID} Components of the Income Process in the PSID}
\centering
\includegraphics[width=1\textwidth]{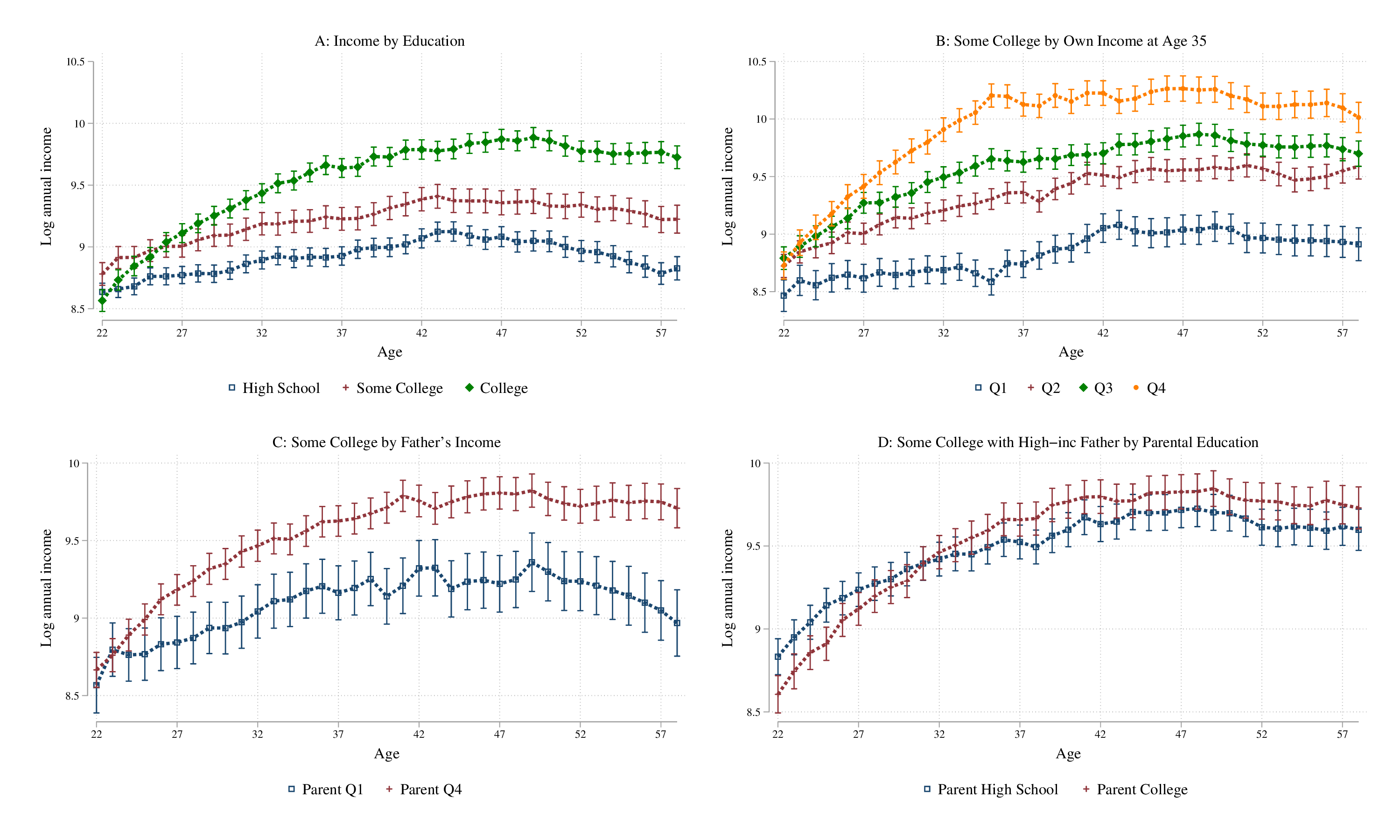}
\fnote{\textbf{Notes:} Panel A shows income trajectories by education category. Panel B focuses only on individuals with at least some college, who are split into four groups according to their annual income at age 35. Category Q1 refers to the bottom and Q4 to the top quartile. In Panel C, individuals with at least some college are divided into two groups, those in the top and those in the bottom quartile of fathers' lifetime income. In Panel D, individuals with at least some college whose parental income belongs to the top median of the distribution are divided into parents with at least some college and parents with only primary schooling. Confidence intervals (95\%) are plotted around each line.}
\end{figure}

We show evidence of each of these components in the Swedish (Figure \ref{fig:fig1}) and the US data (Figure \ref{fig:incomeprocPSID}). Panel A of Figure \ref{fig:fig1} illustrates that, unsurprisingly, income profiles vary with own education (and by occupation, see Appendix Figure A.1). Accounting for observable heterogeneity is, therefore, important. Panel B illustrates that income levels and subsequent income growth are negatively correlated. Splitting college-educated individuals into four quartiles of their annual income at age 35, those in the bottom (top) quartile have the strongest (weakest) income growth in the following years. Transitory shocks are one explanation for this regression to the mean. Apart from attenuating intergenerational estimates via its effects on parental income (\citealt{Atkinson1980}), it may also complicate corrections for lifecycle dynamics in child income (see Section \ref{sec:correctionmethods}).

Panel C provides evidence on a more controversial question, namely if residual income $y^i_{h,t}$ grows at an individual-specific and deterministic rate or follows a random walk. The figure plots the average income profiles of college graduates by the quartile of their \textit{father's} income. Even conditional on education, we observe substantial differences in income growth. College-educated sons with fathers in the top quartile tend to have lower incomes in their mid 20s, but around 40 percent higher incomes around age 40, compared to college-educated sons with fathers in the bottom quartile. We find similar evidence in the PSID (Figure \ref{fig:incomeprocPSID}, Panel C) and when conditioning on occupation (Figure A.1, Panel C) or additional dimensions of family background. For example, in Panel D we show that among college graduates whose father's income is above the median, those with more educated fathers have steeper income profiles than those with less educated fathers (a pattern that replicates in the PSID, see Figure \ref{fig:incomeprocPSID} Panel D; and see \citealt{Michele2018} for related evidence in Italian data).

\begin{table}[h!]
  \centering
  \caption{Heterogeneity in Income Growth by Parental Income (Swedish data) }
   \begin{adjustbox}{width=\textwidth}
    \begin{tabular}{lcccccc}
    \toprule
    \toprule
          & (1)   & (2)   & (3)   & (4)   & (5)   & (6) \\
    \midrule
    Log (Father's Income)/100  &       &       &       &       &       &  \\
    x Age 25-30 & 9.119*** & 3.778*** & 3.421*** & 3.656*** & 1.890*** & 1.887*** \\
          & (0.229) & (0.217) & (0.250) & (0.219) & (0.215) & (0.249) \\
    x Age 30-35 & 4.799*** & 2.034*** & 1.404*** & 2.689*** & 1.651*** & 1.167*** \\
          & (0.194) & (0.198) & (0.225) & (0.197) & (0.198) & (0.227) \\
    x Age 35-40 & 1.276*** & 0.352 & 0.159 & 0.154 & 0.018 & -0.059 \\
          & (0.189) & (0.194) & (0.223) & (0.196) & (0.197) & (0.227) \\
    x Age 40-45 & 0.123 & -0.416* & -0.284 & 0.034 & -0.238 & -0.159 \\
          & (0.177) & (0.183) & (0.210) & (0.184) & (0.186) & (0.214) \\
    x Age 45-50 & -0.223 & 0.118 & -0.015 & 0.154 & 0.238 & 0.021 \\
          & (0.173) & (0.178) & (0.207) & (0.180) & (0.182) & (0.212) \\
    x Age 50-55 & -1.276*** & -0.726*** & -0.560** & -0.663*** & -0.456* & -0.292 \\
          & (0.171) & (0.176) & (0.203) & (0.178) & (0.180) & (0.209) \\
    \midrule
    Education x Age &       & X     & X     &       & X     & X \\
    Occupation x Age &       &       &       & X     & X     & X \\
    Skill scores x Age &       &       & X     &       &       & X \\
    Demographics x Age &       &       &       &       &       & X \\
    \midrule
    N     & 946,534 & 946,534 & 741,467 & 916,201 & 916,201 & 717,582 \\
    $R^2$  & 0.072 & 0.102 & 0.107 & 0.102 & 0.117 & 0.122 \\
    \bottomrule
    \bottomrule
    \end{tabular}%
  \end{adjustbox}
  \smallskip 
     \fnote{\textbf{Notes:} The dependent variable is the five-year change in log annual income (i.e., the difference between log income at age $t$ and log income at $t-5$), measured at $t=30, 35, 40, 45, 50, 55$. Education distinguishes 15 levels of highest educational attainment. Occupation is at the two-digit level (56 groups). Skill scores are cognitive and non-cognitive skill scores from the military draft. Demographic variables are birth order, family size, and an immigrant dummy. All these variables, as well as father's log lifetime income/100, are interacted with the indicators for the six age groups. We remove time effects to abstract from the business cycle. Annual incomes below 20\% of the yearly in-sample median are excluded. Robust standard errors in parentheses, * $p<0.05$, ** $p<0.01$, *** $p<0.001$.}
\label{tab:tabsec3} 
\end{table}%

Could controlling for a broader set of characteristics capture this ``unobserved'' heterogeneity? Table \ref{tab:tabsec3} reports a more systematic analysis, regressing income \textit{growth} on father's income and various background characteristics, all interacted with six different age groups. The dependent variable is the five-year change in log annual income, measured in adjacent intervals, and starting with the difference in log income between age 30 and age 25 in the first row. 

Column (1) reports the raw differences, showing that a log-unit increase in father's income is associated with a 9.1 log point higher income growth between age 25 and 30. This difference in growth rates diminishes over age, and eventually turns negative. The pattern weakens but still holds when controlling for observable characteristics. As shown in column (2), conditional on education the incomes of those with fathers with a log-unit higher income grow more than 6 log points faster between age 25 and 40 but about 1 log point slower between age 40 and 55. These patterns are more pronounced among more highly-educated children, and remain similar when adding controls for cognitive and non-cognitive skill scores from the military draft in column (3). The parental income gradient remains large within 2-digit occupations, but shrinks  when conditioning on both education and occupation (columns 4 and 5). In column (6), we include all controls jointly and add demographic characteristics (birth order, family size, immigrant status). While explaining much of the heterogeneity, income growth in the late 20s and early 30s still increases in parental income. 

Our primary question is whether income growth varies with the \textit{level} of parents' income, but it might also vary with other properties of the parental income process. For example, \cite{halvorsen2021earnings} show that fathers' and children's income growth are correlated even after controlling for fathers’ income and wealth, while \cite{lochner2022earnings} note that conditional on the father's expected lifetime earnings, his earnings \textit{trajectory} remains informative about the child's earnings. While we do not have information on wealth, we confirm in Appendix Table A.2 that income growth patterns between parents and children are correlated early in life (between ages 25 and 30), even after controlling for the father's lifetime income.\footnote{We conduct this test using birth cohorts 1970-78, for whom we observe fathers' income from ages 25 to 40.} This relationship becomes weaker at older ages and when controlling for education, illustrating that this similarity in income trajectories is partly due to correlated educational choices within the family. 

The role of unobserved heterogeneity in income processes remains controversial, as it is difficult to distinguish from stochastic processes with high persistence. By combining long income series with information on family background, one can however provide evidence on this question: income growth varies systematically with parental characteristics, conditional on an individual's own observable characteristics. Because parental characteristics are predetermined, and potentially observed by the child, this pattern is more readily interpreted as a deterministic factor rather than a stochastic shock.  
But while \cite{Guvenen2009} assumes that the individual-specific component of income growth is linear in experience, our evidence suggests that differences by family background matter primarily at young age, and that this association may flip sign at older ages. More generally, differences in skill and earnings growth  decline with age (\citealt{lochner2022earnings}). Such non-linear patterns would be difficult to detect in the higher-order autocovariances of earnings that are often used to identify HIP components (\citealt{Guvenen2009}, \citealt{Hoffmann2019}).

While we do not attempt to identify \textit{why} children from affluent families have steeper income profiles, there are a number of plausible mechanisms. First, human capital investments depend on parental background, either directly, because of credit constraints, or indirectly, because of the effect of wealth on risk aversion (\citealt{blandendoepkestuhlerHB2022}). This may not just affect investments in formal education, but also human capital investments in the early career. Second, differences in the \textit{returns} to human capital investments affect the slope of age-income profiles (\citealt{benporath1967}), and those returns might increase with parental income. Third, children from affluent families tend to find jobs in better-paying firms (\citealt{DobbinZohar}), which may also offer more on-the-job learning or higher returns to experience (\citealt{arellano2021differences,ForsbergNybomStuhler}). The observation that children from affluent families tend to have steeper income profiles matters for distributional questions and may lead to biased estimates of income mobility, as we show next. 

\section{Bias Corrections in the Intergenerational Literature}  \label{sec:correctionmethods}

In many intergenerational studies, the outcome of interest is an individual's lifetime income, yet typically only short snapshots of income are available. To address this issue the literature has considered two alternative approaches: (i) errors-in-variables models that formalize the relation between (observed) annual and (unobserved) lifetime income, and (ii) models of the income process itself, which then determine the relation between annual and lifetime incomes. 

\subsection{Errors-in-Variables Models}

Errors-in-variables models have a long tradition in intergenerational research. The income process is not explicitly modeled, but its assumed properties inform the errors-in-variables assumptions. Many applications rely on a generalized errors-in-variables model proposed by \cite{Haider2006}, which allows for the relation between annual and lifetime incomes to vary systematically over the lifecycle. It implies that lifecycle bias is reduced by measuring incomes in midlife, a simple rule-of-thumb that has become widely adopted. While indeed reducing bias, this strategy is subject to some limitations (see also Appendix B). First, income growth varies with parental income even conditional on own lifetime income, such that the bias may not be fully eliminated at the ``optimal'' age as prescribed by the model \citep{Nybom2016}. More importantly, this optimal age may vary across countries or time but is typically unknown in applications. Researchers therefore measure incomes at \textit{some} age in midlife, subject to data limitations, resulting in substantial age differences across studies (see Table A.1). As even slight age variations affect the IGE (Table B.1), existing estimates are difficult to compare. Finally, prime-age incomes may simply not be observable for the population of interest, such as for recent birth cohorts.

\subsection{Modelling the Income Process}\label{subsec:modellingprocess}

An alternative is to model the income process directly. Age-income profiles are first estimated based on partial income profiles and individual characteristics. These first-step estimates are then used to predict lifetime income for each person. The main challenge is to extrapolate (observed) income spans to the complete lifecycle without inducing biases that co-vary with explanatory variables of interest (e.g., parental income). We propose such a ``lifecycle'' estimator in Section \ref{sec:lifecyclestimator}. To motivate our approach, however, it is instructive to first review existing work on this problem. 

\textit{Accounting for Individual Characteristics.}
One strategy is to model the income process as a function of an individual's own characteristics. In a first step, we estimate
\begin{equation} \label{eqvogelhertz}
y_{ict} = \alpha_i + g(A_{ict},Z_{ic}) + \varepsilon_{ict}, 
\end{equation}
where $y_{ict}$ is log income of individual $i$ from cohort $c$ in year $t$, $\alpha_i$ an individual fixed effect, and $g(A_{ict},Z_{ic})$ an interaction of age and a vector of individual characteristics $Z_{ic}$ (e.g. education). The estimates can then be used to predict lifetime incomes. Different approaches are used to address the issue that income profiles are only partially observed. \cite{Hertz2007} predicts incomes at a given age, whereas \cite{Vogel2007} predicts the entire lifecycle under the assumption that parents and children have similar income profiles (conditional on covariates). \cite{Justman2021} pool many cohorts to estimate lifecycle profiles, an approach that we follow in our trends analysis in Section \ref{sec:correctiontrends}.

By allowing for income growth to differ by education, an important source of heterogeneity can be accounted for (Panel A in Figures \ref{fig:fig1} and \ref{fig:incomeprocPSID}). However, our finding that income growth varies with parental income even within education or occupation groups (Panel C in Figures \ref{fig:fig1}, \ref{fig:incomeprocPSID} and A.1) suggests that the procedure may remain sensitive to lifecycle effects. Appendix Table A.3 probes this hypothesis by comparing estimates from partial profiles against the ``true'' IGE based on lifetime incomes (for Sweden). We first estimate equation (\ref{eqvogelhertz}) using incomes from a given age range and flexible age-education interactions to predict incomes at a given age (Panel A). While much better than directly using annual income at that same age (bottom panel), the corrected estimates still increase with (i) the age range included in the estimation and (ii) the age at which incomes are predicted, even if aggregating predictions over the entire lifecycle (Panel B).\footnote{Specifically, we split each individual's income profile into two halves, with income for the ``younger'' copy assumed to be observed in each of the age ranges of Table A.3, and the ``older'' copy being observed thereafter. This allows us to focus on the problem of missing income information for a given person, while abstracting from the issue that certain age ranges are missing for the entire population of interest. We follow the same strategy in Section \ref{sec:newcorrection}.}

 Why do estimates from equation (\ref{eqvogelhertz}) remain volatile? Because children from high-income families experience higher income growth even given own education or occupation (Section \ref{sec:incprocess}), estimates of the fixed effects $\alpha_i$ – and thus lifetime incomes – depend on the observed age range. For example, when observing only early (late) ages, we understate (overstate) the lifetime income of those with low initial incomes but stronger growth. The earlier incomes are observed, the more the IGE is understated (see also Appendix C). This issue also affects estimation of mobility \textit{trends}. Many studies keep the age at which incomes are predicted fixed over cohorts, thereby eliminating the variability across the columns in Panel A of Table A.3. But IGE estimates also vary with the sample's age composition, across the rows of Table A.3. For example, predicting incomes at age 25, the estimates shift from 0.042 to 0.147 when the sample extends from age 25-30 to age 25-45. This age composition is typically not held constant when estimating mobility trends.

An alternative approach allows for income growth to vary with income \textit{levels}.\footnote{While such heterogeneity could be captured by estimating individual-specific slopes (as in \citealt{Jantti2012}), we do not pursue this option here as direct extrapolation from partially observed slopes would produce unstable predictions of lifetime income if only few income observations are available per person (see also \citealt{Jenkins2009}).} This approach is rarely used in applications, but its potential advantages have been studied by \cite{Creedy1988}. To understand the basic argument, assume that individuals maintain a constant \textit{relative position} in the income distribution. Researchers can then combine relative positions with information on the income distribution at each age to construct lifetime incomes. Appendix D provides details on this approach, showing that it tends to overstate the IGE. This bias stems from mean reversion (see Figure \ref{fig:fig1}C): due to transitory shocks, income growth and levels are negatively correlated. Still, the observation that income levels and growth are correlated is useful and we revisit it below.

\section{A Lifecycle Estimator for the Intergenerational Elasticity}  \label{sec:newcorrection}

In this section, we propose a lifecycle estimator of intergenerational mobility in lifetime income. The key assumption for unbiased estimation of the IGE is that the prediction error in children’s lifetime income is uncorrelated with parental income. Accordingly, our proposed estimator controls for the relation between parental income and the shape of age-income profiles of their children. Using Swedish and US data, we illustrate that the estimator can be applied in diverse settings and that it provides more robust estimates of the IGE than other approaches.

\subsection{Econometric Specifications} \label{sec:lifecyclestimator}

The estimation consists of two steps. In a first step, we estimate and predict the individual lifecycle income profiles in the child generation, based on partial income snapshots and individual and parental characteristics. In a second step, we estimate the IGE using lifetime incomes based on the predicted profiles. This lifecycle estimator uses the available income information more fully than the commonly used rule-of-thumb approaches based on income averages. Our two-step approach builds on earlier contributions (see Section \ref{sec:correctionmethods}), but explicitly accounts for income growth to vary with parental income even conditional on own characteristics (see Section \ref{sec:incprocess}).

Specifically, in the first step we use OLS to estimate variants of
\begin{equation} \label{eqtwostep}
y_{ict} = \alpha_i + g(A_{ict},Z_{ic}) + f(A_{ict},Z_{ic},P_{ic}) + \varepsilon_{ict}, 
\end{equation}
where $y_{ict}$ is log income of individual $i$ from cohort $c$ in year $t$, $\alpha_i$ are individual fixed effects, $g(A_{ict},Z_{ic})$ represents interactions between age and a vector $Z_{ic}$ of the individual's own characteristics (e.g., education), and $f(A_{ict},Z_{ic},P_{ic})$ interacts age, education, and parental characteristics $P_{ic}$. In our application, $P_{ic}$ contains log parental income and indicators for parental education. Our preferred specification allows for a quadratic in age in $f(\cdot)$, as income growth varies more strongly with parental background in the early career than at later ages (see Figures \ref{fig:fig1} and \ref{fig:incomeprocPSID}). We also consider two other variants of this estimator:

\textbf{No-FE estimator.} We estimate equation \eqref{eqtwostep} with or without individual fixed effects $\alpha_i$. While allowing for individual intercepts might seem an obvious improvement, the flexibility of a full set of fixed effects comes at a cost, making it harder to capture the heterogeneity in income \textit{slopes} -- especially when only short snapshots of incomes are observed and $f(A_{ict},Z_{ic},P_{ic})$ may be misspecified. In such cases, it may be preferable to allow intercepts to vary only with the regressor of interest (i.e., replacing $\alpha_i$ with a function of parental income), as we demonstrate below.

\textbf{Slope-level estimator.} If parental income is not well observed, one may instead allow income slopes to vary with the level of an individual's \textit{own} income. This \textit{slope-level} estimator is motivated by the fact that those from affluent families have both higher levels and steeper slopes compared to others (Figures \ref{fig:fig1} and \ref{fig:incomeprocPSID}, Panel C and Figure A.2), implying a positive relation between income levels and growth (\citealt{Creedy1988}). We allow for income growth to vary with the individual fixed effect rather than current income, to address the mean reversion in the latter due to transitory noise (see Section \ref{sec:incprocess}). Specifically, we estimate 
\begin{equation} \label{eqtwostepFE}
y_{ict} = \mu_i + g(A_{ict},Z_{ic}) + f(A_{ict},Z_{ic},\mu_i) + \nu_{ict}, 
\end{equation}
where individual characteristics $Z_{ic}$ are interacted with age and the individual fixed effect $\mu_i$. This model can be estimated recursively (as in \citealt{Roca2017}).\footnote{We first approximate the individual fixed effect $\mu_i$ by estimating equation (\ref{eqtwostepFE}) while omitting $f(\cdot)$. We then estimate the complete equation (\ref{eqtwostepFE}) with $\widehat{\mu}_i$ included in $f(\cdot)$. This second step can be iterated until estimates of the fixed effects converge, but as further iterations have only negligible effects we report estimates from a single iteration below.} Our preferred implementation interacts $\mu_i$ with a quadratic in age, since income growth varies more at early age.

All three first-step models above are parametric, in which we specify the models at our own discretion. Alternatively, one could consider data-driven approaches to select the first-step specification. We motivate our focus on parametric models by the fact that we know the main source of bias -- the relationship between parental income and own earnings growth -- but also explore the performance of more data-driven methods in Section \ref{sec:perf_sweden}.

After the first-step estimation of equation (\ref{eqtwostep}) or (\ref{eqtwostepFE}), we predict log annual incomes for each person between age 25-58 for Sweden and age 22-58 for the US, and then convert them into absolute incomes. We then construct the individual sum of all annual incomes to, finally, estimate the typical IGE regression, regressing the log lifetime income of the child on the log income of the parent.\footnote{Taking the average instead of the sum of annual incomes would not affect our estimates, as the denominator (\# of annual income observations) is constant across individuals and would only enter the intercept.} 

However, this approach is subject to three conceptual issues. First, the estimation consists of multiple steps, which affect statistical inference. As our benchmark sample for Sweden is large, we initially ignore sampling error in the first-step estimation, but later study how sensitive the estimators are to sample size. Second, the dependent variable in equation (\ref{eqtwostep}) is the \textit{log} of annual income, and conversion to absolute incomes for the construction of lifetime incomes gives rise to a well-known re-transformation problem.\footnote{While the fitted values from the estimation of equation (\ref{eqtwostep}) have mean zero by construction ($E[\hat{\varepsilon}_{ict}]=0$), their mean will be positive after transformation ($E[exp(\hat{\varepsilon}_{ict})]>0$). If this expectation were constant across individuals and linearly separable in log lifetime income, it would only affect the intercept of the intergenerational regression, not its slope. But $E[exp(\hat{\varepsilon}_{ict})]$ tends to be larger for individuals with low lifetime income if their income is more variable around the mean tendency over the lifecycle.} We address this issue using the solution proposed by \cite{Wooldridge2006}.\footnote{Specifically, we estimate complete lifecycle profiles of each individual in the child generation, based on a quartic in age interacted with education dummies and individual fixed effects, to construct $SM_{ic}=\sum_{t=25}^{58} exp(\hat{\varepsilon}_{ict})$ to adjust the predicted lifetime income accordingly.} Alternatively, we estimate a version of equation (\ref{eqtwostep}) in which $y_{ict}$ is incomes in levels instead of logs (see Appendix E). Third, in many applications the population of interest is only observed for a certain age range (e.g. at young age), so their income profiles need to be extrapolated over the unobserved age range. We initially abstract from this issue by exploiting that our benchmark samples include long income series for each person. Specifically, we randomly split each individual's income profile into two parts, a ``young'' or an ``old''. Each part is then assumed to be a different individual by receiving separate identifiers.\footnote{We proceed with this random split of the sample into two parts for our analysis using the Swedish data. Due to its small sample size, for the PSID analysis, we duplicate each individual's income profile into two parts, a ``young'' or an ``old''. Each copy assumes a different individual identifier.} If assigned to the ``young'' group, the income profile is assumed to be observed only up to some age threshold, while if assigned to the ``old'' group the income profile is assumed to be observed only thereafter. This allows us to focus on the problem of missing incomes for a given person, while abstracting from the issue that certain age ranges may be missing for the entire population of interest. We return to this extrapolation issue in a robustness analysis, and in Section \ref{sec:correctiontrends} when estimating mobility trends for recent cohorts.

\subsection{Performance of the Lifecycle Estimator in Swedish Registers}\label{sec:perf_sweden}

\singlespacing
\begin{table}
\centering
  \caption{The Lifecycle Estimator in the Swedish Data}
     \begin{adjustbox}{width=\textwidth}
         \begin{tabular}{lcccccccc}
    \toprule
    \toprule
          & \multicolumn{2}{c}{Direct estimator} &       & \multicolumn{5}{c}{Lifecycle estimator} \\
\cmidrule{2-3}\cmidrule{5-9}          & Lifetime & Annual &       & Baseline  & Parental & Parental & Parental & Slope-level \\
          &       &       &       &       & Linear & Quadratic & Quadratic & Quadratic \\
              &       &       &       &   FE  & FE & FE & no FE & FE \\
    Son's Age & (1)   & (2)   &       & (3)   & (4)   & (5)   & (6)   & (7) \\
    \midrule
    Age 25-27 & 0.253 & 0.046 &       & 0.150 & 0.199 & 0.236 & 0.266 & 0.195 \\
    N=94,100 & (0.004) & (0.002) &       & (0.004) & (0.004) & (0.004) & (0.002) & (0.006) \\
    $R^2$ 1st step & -     & -     &       & 0.193 & 0.198 & 0.202 & 0.006 & 0.221 \\
    $R^2$ 2nd step & 0.050 & 0.001 &       & 0.017 & 0.030 & 0.042 & 0.202 & 0.013 \\
          &       &       &       &       &       &       &       &  \\
    Age 25-30 & 0.253 & 0.101 &       & 0.188 & 0.225 & 0.265 & 0.265 & 0.248 \\
    N=94,194 & (0.004) & (0.002) &       & (0.003) & (0.003) & (0.003) & (0.002) & (0.005) \\
    $R^2$ 1st step & -     & -     &       & 0.317 & 0.321 & 0.325 & 0.006 & 0.366 \\
    $R^2$ 2nd step & 0.050 & 0.006 &       & 0.030 & 0.043 & 0.059 & 0.201 & 0.026 \\
          &       &       &       &       &       &       &       &  \\
    Age 25-35 & 0.254 & 0.159 &       & 0.206 & 0.231 & 0.263 & 0.263 & 0.254 \\
    N=94,264 & (0.004) & (0.001) &       & (0.003) & (0.003) & (0.003) & (0.002) & (0.004) \\
    $R^2$ 1st step & -     & -     &       & 0.485 & 0.487 & 0.489 & 0.006 & 0.536 \\
    $R^2$ 2nd step & 0.050 & 0.013 &       & 0.040 & 0.050 & 0.064 & 0.200 & 0.036 \\
          &       &       &       &       &       &       &       &  \\
    Age 25-40 & 0.254 & 0.204 &       & 0.225 & 0.270 & 0.277 & 0.262 & 0.269 \\
    N=94,311 & (0.004) & (0.001) &       & (0.003) & (0.003) & (0.003) & (0.002) & (0.004) \\
    $R^2$ 1st step & -     & -     &       & 0.646 & 0.648 & 0.648 & 0.006 & 0.678 \\
    $R^2$ 2nd step & 0.050 & 0.018 &       & 0.046 & 0.066 & 0.069 & 0.198 & 0.042 \\
          &       &       &       &       &       &       &       &  \\
    Age 25-45 & 0.254 & 0.234 &       & 0.239 & 0.286 & 0.274 & 0.261 & 0.272 \\
    N=94,339 & (0.004) & (0.001) &       & (0.003) & (0.003) & (0.003) & (0.002) & (0.004) \\
    $R^2$ 1st step & -     & -     &       & 0.783 & 0.783 & 0.783 & 0.006 & 0.800 \\
    $R^2$ 2nd step & 0.050 & 0.021 &       & 0.051 & 0.071 & 0.066 & 0.197 & 0.047 \\
    \bottomrule
    \bottomrule
    \end{tabular}%
    \end{adjustbox}
    \smallskip
  \fnote{\textbf{Notes:} The table reports the slope coefficient from a regression of son's income on father's lifetime income. The measure for son's income is log lifetime income in column (1), the pooled log annual incomes from age 25 to the indicated upper age bound in column (2), or the predicted lifetime income from a first-step estimation in the indicated age range of equation \eqref{eqtwostep} in columns (3)-(6) or equation \eqref{eqtwostepFE} in column (7). See text for detailed definitions of each estimator.``$R^2$ 1st step'' is the $R^2$ from a regression of son's actual lifetime incomes on predicted lifetime incomes based on the observed age range. ``$R^2$ 2nd step'' is the $R^2$ from the regression of the predicted log lifetime income of sons on the lifetime income of fathers. Robust standard errors in parentheses.}
  \label{tab:lifecycleestimator}%
\end{table}%
\doublespacing

Table \ref{tab:lifecycleestimator} presents evidence on the performance of the proposed lifecycle estimator. We consider different age thresholds, assuming that child income is observed only over age 25-27 (first panel), age 25-30 (second panel), and so on. Column (1) reports benchmark estimates of the IGE based on ``true'' lifetime incomes, which are about $\widehat{\beta}=0.25$. 
In column (2), we report estimates based on pooled annual incomes from age 25 to the indicated upper age bound (e.g. age 25-27 in the first row). Consistent with prior evidence on lifecycle bias, those estimates are very sensitive to the age range, being as low as $0.05$ when child incomes are measured only until age 27 and increasing monotonically when increasing that age range. 

\begin{figure}[tb]
 \centering
 \caption{\label{fig:parental_actual_pred}{Comparison between Actual and Predicted Profiles}}
   \includegraphics[width=0.8\textwidth]{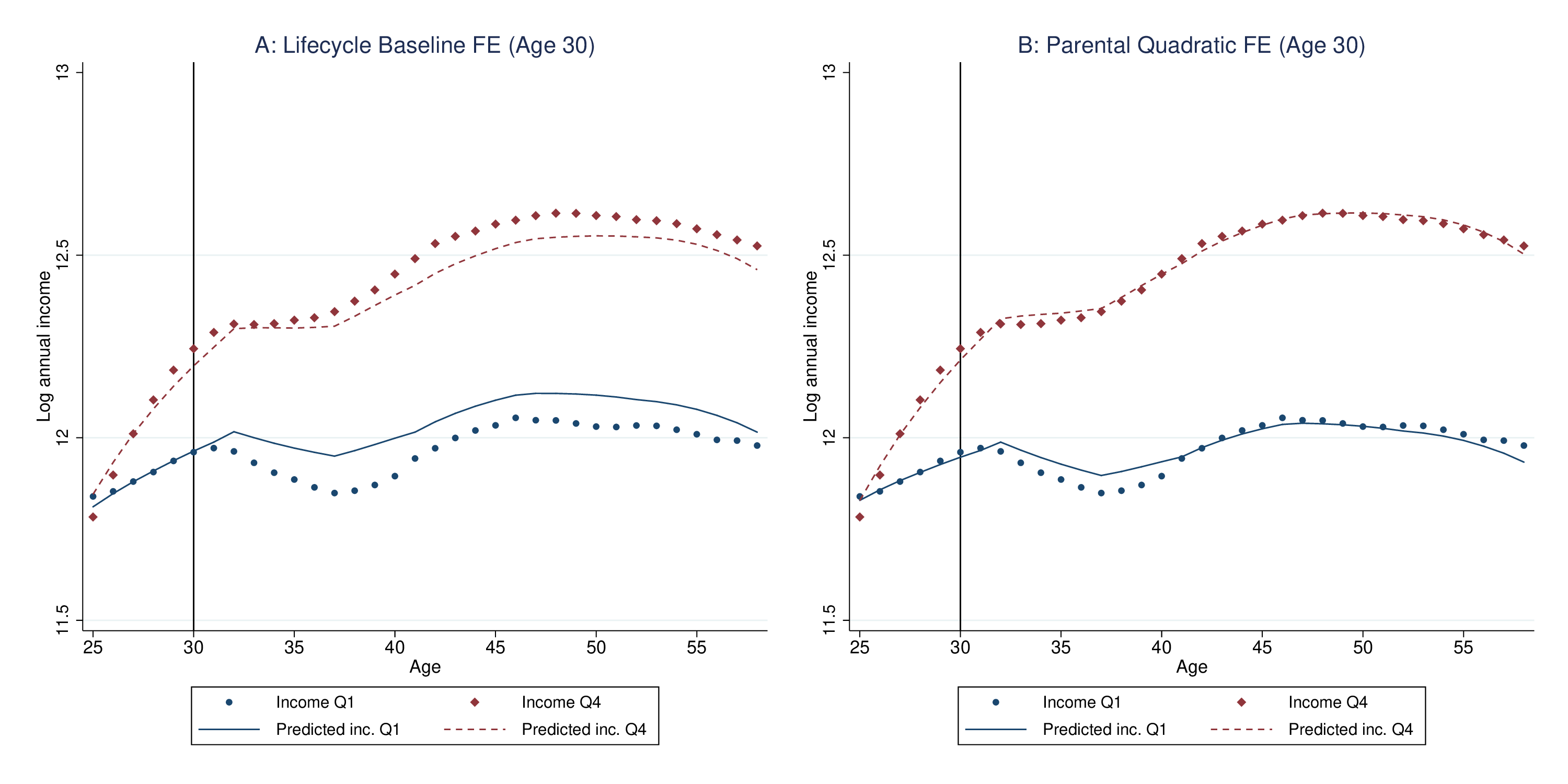}
    \fnote{\textbf{Notes:} The figure plots the actual log income profiles and predicted profiles separately for the top and the bottom quartiles of parental income. To predict complete profiles based on income observations until age 30 we implement a lifecycle estimator without parental interactions (baseline, Panel A) or with quadratic interactions between child age and parental income (parental quadratic, Panel B).}
\end{figure}

In columns (3) to (7) we implement different variants of the lifecycle estimator. Column (3) reports estimates from a baseline estimator based on equation (\ref{eqtwostep}) that distinguishes four education groups (as defined in Figure A.3), but does not include parental characteristics $P_{ic}$. This estimator is similar in spirit to those used in prior studies (see Section \ref{sec:correctionmethods}), and performs better than a direct estimator using annual incomes. However, it still varies with the age at which incomes are measured and understates the IGE by nearly one third when child incomes are measured at age 25-30. As discussed in Section \ref{sec:correctionmethods}, this estimator remains sensitive to age since it does not account for differences in income growth by parental background. This issue is also illustrated in Panel A of Figure \ref{fig:parental_actual_pred}, which plots the mean actual (dots) and predicted (solid lines) lifecycle profiles of children from the top (red) or bottom quarter (blue) of the parental income distribution when child income is observed at age 25-30. The baseline lifecycle estimator understates the true income growth at the top and overstates growth at the bottom of the parental income distribution, leading to downward-biased estimates of the IGE. 

Column (4) therefore reports estimates from the ``parental'' lifecycle estimator, in which the first-step estimation of equation (\ref{eqtwostep}) includes linear interactions between child age and parental income and education. This estimator performs better than the baseline estimator, in particular at young ages, as it captures some of the heterogeneity in income growth by parental background. However, the estimated IGE still increases systematically with the age at which incomes are measured, and understates the true IGE when incomes are measured at early age. The reason follows from Figure \ref{fig:parental_actual_pred}: the association between income growth and parental background is more pronounced at early than late ages, so linear extrapolations from early age work poorly. 

We, therefore, consider a ``parental quadratic'' lifecycle estimator in column (5), which uses a quadratic rather than linear polynomial in child age interacted with parental income. The estimates are now close to the benchmark for all age ranges, reflecting that quadratic interactions capture the heterogeneity in income slopes by parental background well, as also illustrated in Figure \ref{fig:parental_actual_pred}B. However, even with quadratic interactions, some bias remains when measuring incomes in only a short range at young age (e.g., age 25-27). This bias reflects a form of overfitting that can occur due to the high flexibility of a model with individual fixed effects. Specifically, if the functional form of $f(A_{ict},Z_{ic},P_{ic})$ in equation (\ref{eqtwostep}) does not correspond to the true functional form, some of the heterogeneity in income growth may instead be captured by the individual fixed effects $\alpha_i$.\footnote{While the ``parental quadratic'' estimator includes a quadratic in child age interacted with parental income, we observe that children from high-income parents experience even steeper income growth in the first years of their career than would be captured by this quadratic interaction. As a consequence, when splitting the income profiles of children into a ``young'' and ``old'' copy to estimate equation \eqref{eqtwostep}, we estimate larger fixed effects for the ``old'' than the ``young'' part. Since a time-constant fixed effect cannot provide a good approximation for differences in income growth over many years, this issue arises only if child age is observed in a very narrow age range.} 

Column (6) therefore shows estimates from a lifecycle estimator without individual fixed effects (``no-FE''), in which intercepts can vary with parental income but individual variation around that mean tendency is disregarded. This alternative estimator is insensitive to the age at which incomes are observed, and is always close to the benchmark. It captures heterogeneity in income slopes even if incomes are observed only at a very young age (age 25-27). These results illustrate that if the object of interest is income differences by parental background, it can be advantageous to model only that specific form of heterogeneity rather than individual-level variation in the intercepts. 

Finally, column (7) shows estimates from the alternative ``slope-level'' estimator (see Section \ref{subsec:modellingprocess}) based on equation (\ref{eqtwostepFE}), which interacts a quadratic in age with the individual's own (estimated) individual fixed effect rather than parental income. The estimator performs largely similarly to the corresponding parental lifecycle estimator, suggesting that systematic variation in income growth by parental background could potentially be addressed indirectly, without observing parental characteristics, by accounting for the covariance between income levels and growth. However, this estimator performs worse when incomes are observed only at age 25-27, presumably because income levels are not very predictive about long-term income at such early age. 

\textbf{Robustness.} The estimates in Table \ref{tab:lifecycleestimator} are based on large samples with many income observations. In many applications, however, researchers observe fewer individuals or fewer income observations per individual. We therefore test our lifecycle estimator(s) in such settings. We first study how its performance varies with the number of incomes available per child. Specifically, we randomly select six annual incomes per person within the indicated age range and successively drop income observations until only two remain per person. Table A.4 shows that while the noise in the estimation of lifetime incomes increases and the $R^2$ in the IGE regression drops, the mean of the estimates remains stable.

A second concern is that the shape of lifecycle profiles cannot be precisely estimated in smaller samples. To probe this concern, Table A.5 reports estimates from differently sized samples. We draw fractions $1/k$ of our original sample (as indicated in the top row) and then implement the benchmark estimator based on lifetime incomes, as well as the \textit {parental} (with individual fixed effects) and \textit{slope-level}  lifecycle estimators. The table reports the mean and standard deviation of the IGE estimates across repeated draws from the main sample. The mean appears robust to sample size, and while the precision of the estimators decreases in smaller samples, so do the corresponding benchmark estimates based on observed lifetime incomes. 

A third concern is that our estimator(s) may work well because we use the same cohorts and income years in both estimation steps, but that they would perform worse in less ideal situations. We thus explore their out-of-sample performance  by varying the cohorts and income years used in the first-step prediction, and then using these predictions to estimate the IGE for our baseline cohorts born 1952-1960.\footnote{Note that we here use the trends sample (see next section), which implies that the corresponding benchmark estimate is slightly lower.} This exercise approximates the type of settings in which our estimators will be particularly useful. Table A.6 shows in column (2) that the estimators are largely unaffected when including a wider set of cohorts in the first step. However, columns (3) and (4) indicate that the estimators overstate the IGE somewhat when we in addition only consider more recent income years, such that we observe the baseline cohorts only at a relatively old age. Still, this upward bias tends to be smaller compared to when directly using all observed incomes (see row 2).

\textbf{Performance of the First-Step Estimation.} While our objective is to reduce bias in estimates of the IGE, an interesting question is to what extent controlling for parental income can improve the lifetime-income predictions in the first estimation step. Separately for each estimator and age interval in which child incomes are observed, Table \ref{tab:lifecycleestimator} reports the $R^2$ in a regression of the children's actual on their predicted \textit{lifetime} income ($R^2$ 1st step). Naturally, the $R^2$ is low if only short snapshots of income are observed but increases when incomes are observed over longer age ranges (compare rows). For a given age range, the $R^2$ is slightly higher when allowing the income slopes to vary with parental income (cf. columns 3 and 4), but the gains are small. Despite these small gains in predictive power, the correction strongly improves estimates of the IGE. The reason is that even variability that is negligible in an $R^2$ sense can have strong effects on IGE estimates if it is directly related to parental income. Our approach is therefore superior not so much in terms of overall explanatory power, but in the sense of \textit{reducing systematic bias} in mobility estimates. 

\textbf{ML/Regularization.} An alternative to our parametric estimator is to use machine learning (ML) methods, such as lasso or elastic nets, to select the first-step predictors. However, such plug-in ML methods would optimize predictive accuracy in the first-step estimation of income profiles (by balancing bias and variance), whereas our objective is to reduce bias in the second-step estimation of the IGE. Although the two objectives are related, the key source of bias in the IGE –– variation in income growth rates by parental income (see Figure \ref{fig:fig1}) -- may not be a particularly strong determinant of lifetime incomes, as noted in the previous section.

\singlespacing
\begin{table}
\centering
  \caption{ML Estimation of Lifecycle Profiles}
     \begin{adjustbox}{width=\textwidth}
         \begin{tabular}{lcccccccc}
    \toprule
    \toprule
          & \multicolumn{1}{c}{Direct} &       & \multicolumn{6}{c}{Lifecycle estimator} \\
\cmidrule{2-2}\cmidrule{4-9}          & Lifetime &       & Parental & Lasso & Lasso & Lasso & Lasso & Lasso \\
          &       &       &  Quadratic & 
          &  &  &  & (not pen.) \\          
    Son's Age & (1)  & & (2)  & (3)   & (4)   & (5)   & (6)   & (7) \\
    \midrule
Age 25-27 & 0.219 &  & 0.203 & 0.040 & 0.117 & 0.173 & 0.203 & 0.203 \\
N=71,794 & (0.003) &  & (0.004) & (0.004) & (0.004) & (0.004) & (0.004) & (0.004) \\
$\lambda$ &  &  &  & .01 & .001 & .0001 & .00001 & .01 \\
\# vars &  &  & 76 & 233 & 233 & 233 & 233 & 233 \\
\# vars selected &   &  & 76 & 16 & 48 & 105 & 174 & 42 \\ 
    \bottomrule
    \bottomrule
    \end{tabular}%
    \end{adjustbox}
    \smallskip
  \fnote{\textbf{Notes:} The table reports the slope coefficient from a regression of son’s income on father’s lifetime income. The measure for son’s income is lifetime income in column (1) or the predicted lifetime income from a first-step estimation in the indicated age range of equation (\ref{eqtwostepFE}) in columns (3)-(7). In column (7) we include the parental income x child age interactions as non-penalized regressors in the first step. See text for a description of each estimator. Standard errors in parentheses.}
  \label{tab:ML100p}%
\end{table}%
\doublespacing

We study different ML estimators in Appendix G. Table \ref{tab:ML100p} provides a summary, comparing IGE estimates using our preferred parametric estimator (``parental quadratic'', column 2) with those based on lasso to select the first-step predictors. To show how the latter's performance varies with the number of selected predictors, we also vary the lasso tuning parameter ($\lambda$, columns 3-6). We include a broad range of candidate predictors: alongside our standard variables (e.g. child education, parental income) we also consider family size, birth order, immigrant status, cognitive and non-cognitive skill scores (as in Table \ref{tab:tabsec3}), and all two-way interactions with child age and age squared, resulting in 233 candidate variables. We also include individual FEs in all specifications. As the skill scores contain missings, the ML sample is smaller than our main intergenerational sample used in the rest of the paper, resulting in a slightly smaller benchmark IGE (column 1).\footnote{We confirmed our arguments also in the sample as used in Table \ref{tab:lifecycleestimator}, using a more restricted set of predictor variables. But here we ask whether machine learning methods could offer benefits if a wide set of potential predictors is available.}  

When selecting only a limited set of predictors, the lasso-based estimator performs substantially worse than our preferred parametric estimator. For instance, the IGE estimate is only 0.117 when selecting 48 predictors (column 4), rising to 0.173 with 105 non-zero predictors (column 5), compared to 0.203 with our parametric first step with 76 predictors (column 2). Despite selecting more predictors, the lasso estimator performs worse as it does not reliably select the interactions between child age and parental income that, from Figure \ref{fig:fig1}, we know are crucial sources of bias in the second-step estimation of the IGE. Indeed, when including these interactions as non-penalized regressors, lasso performs similarly to our parametric estimator (column 7). Lasso also performs well when selecting a very small $\lambda$ (column 6), as the resulting long list of predictors then also includes the interactions between age and parental income. In Appendix G, we extend this evidence to other age groups (Table G.1), to using ``postselection'' rather than penalized lasso coefficients in the first step (Table G.2), and to lasso and elastic net estimates in smaller samples with cross-validated tuning parameters (Table G.3). Although the performance of the ML-based estimators varies across specifications, they perform either worse or similar to our parametric first step. 

In sum, plugging standard ML methods into the first-step estimation will in many cases lead to larger biases than our proposed parametric specifications. The issue is that plug-in ML methods do not reliably select the predictors capturing the higher income growth among children from high-income families, which is the key source of bias in the second-step estimation of the IGE. ML methods may still perform well when a broad set of predictors can be selected in large samples, or when including the crucial interactions between child age and parental income as non-penalized regressors. A promising strategy for future research may be to construct a debiased ML estimator for the IGE based on \textit{orthogonal} moments (see \citealt{chernozhukov2022locally, Puerta2024RobustIGE}).

\subsection{Performance of the Lifecycle Estimator in the PSID}

\singlespacing
\begin{table}[t!]
\centering
  \caption{The Lifecycle Estimator in the US Data (PSID)}
     \begin{adjustbox}{width=\textwidth}
         \begin{tabular}{lcccccccc}
    \toprule
    \toprule
          & \multicolumn{2}{c}{Direct estimator} &       & \multicolumn{5}{c}{Lifecycle estimator} \\
\cmidrule{2-3}\cmidrule{5-9}          & Lifetime & Annual &       & Baseline  & Parental & Parental & Parental & Slope-level \\
          &       &       &       &       & Linear & Quadratic & Quadratic & Quadratic \\
              &       &       &       &   FE  & FE & FE & no FE & FE \\
    Son's Age & (1)   & (2)   &       & (3)   & (4)   & (5)   & (6)   & (7) \\
\cmidrule{1-3}\cmidrule{5-9}    Age 22-27 & 0.426 & 0.242 &       & 0.316 & 0.373 & 0.375 & 0.428 & 0.373 \\
    N=892 & (0.035) & (0.016) &       & (0.034) & (0.034) & (0.034) & (0.014) & (0.045) \\
    $R^2$ 1st - lifetime & -     & -     &       & 0.480 & 0.489 & 0.490 & 0.270 & 0.435 \\
    $R^2$ 2nd - lifetime & 0.142 & 0.040 &       & 0.089 & 0.121 & 0.122 & 0.520 & 0.073 \\
          &       &       &       &       &       &       &       &  \\
    Age 22-30 & 0.426 & 0.280 &       & 0.359 & 0.416 & 0.432 & 0.428 & 0.421 \\
    N=892 & (0.035) & (0.013) &       & (0.033) & (0.033) & (0.033) & (0.014) & (0.042) \\
    $R^2$ 1st - lifetime & -     & -     &       & 0.614 & 0.620 & 0.622 & 0.270 & 0.589 \\
    $R^2$ 2nd - lifetime & 0.142 & 0.053 &       & 0.115 & 0.151 & 0.162 & 0.520 & 0.103 \\
          &       &       &       &       &       &       &       &  \\
    Age 22-35 & 0.426 & 0.321 &       & 0.400 & 0.449 & 0.456 & 0.428 & 0.464 \\
    N=892 & (0.035) & (0.011) &       & (0.034) & (0.034) & (0.034) & (0.014) & (0.042) \\
   $R^2$ 1st - lifetime & -     & -     &       & 0.740 & 0.744 & 0.745 & 0.270 & 0.734 \\
    $R^2$ 2nd - lifetime & 0.142 & 0.064 &       & 0.131 & 0.163 & 0.167 & 0.520 & 0.121 \\
          &       &       &       &       &       &       &       &  \\
    Age 22-40 & 0.426 & 0.354 &       & 0.413 & 0.465 & 0.465 & 0.428 & 0.464 \\
    N=892 & (0.035) & (0.010) &       & (0.034) & (0.034) & (0.034) & (0.014) & (0.040) \\
   $R^2$ 1st - lifetime & -     & -     &       & 0.829 & 0.832 & 0.832 & 0.270 & 0.830 \\
    $R^2$ 2nd - lifetime & 0.142 & 0.070 &       & 0.139 & 0.172 & 0.172 & 0.520 & 0.131 \\
          &       &       &       &       &       &       &       &  \\
    Age 22-45 & 0.426 & 0.373 &       & 0.397 & 0.426 & 0.421 & 0.428 & 0.426 \\
    N=892 & (0.035) & (0.009) &       & (0.034) & (0.034) & (0.034) & (0.014) & (0.037) \\
    $R^2$ 1st - lifetime & -     & -     &       & 0.900 & 0.901 & 0.901 & 0.270 & 0.900 \\
    $R^2$ 2nd - lifetime & 0.142 & 0.073 &       & 0.135 & 0.153 & 0.150 & 0.520 & 0.129 \\
     \bottomrule
    \bottomrule
    \end{tabular}%
    \end{adjustbox}
    \smallskip
     \fnote{\textbf{Notes:} The table reports the slope coefficient from a regression of son's income on father's lifetime income. The measure for son's income is log lifetime income in column (1), the pooled log annual incomes from age 25 to the indicated upper age bound in column (2), or the predicted lifetime income from a first-step estimation in the indicated age range of equation \eqref{eqtwostep} in columns (3)-(6) or equation \eqref{eqtwostepFE} in column (7). See text for detailed definitions of each estimator. ``$R^2$ 1st step'' is the $R^2$ from a regression of son's actual lifetime incomes on predicted lifetime incomes based on the observed age range. ``$R^2$ 2nd step'' is the $R^2$ from the regression of the predicted log lifetime income of sons on the lifetime income of fathers. Robust standard errors in parentheses.}
 \label{tab:lifecyclepsid}%
\end{table}%
\doublespacing

To probe the generalizability of these findings, we study how the lifecycle estimator performs using the PSID. We focus on our benchmark sample, born 1952-1960. For these cohorts, we observe nearly complete income profiles and can perform an exercise analogous to the one conducted for Sweden above. We again consider different age thresholds, assuming that child income is observed only at age 22-27, age 22-30, and so on. Column (1) of Table \ref{tab:lifecyclepsid} shows that our benchmark estimate based on ``true'' lifetime incomes for the child generation is around 0.43. This estimate is similar to others reported in the literature, but is still downward biased from the use of noisy incomes in the \textit{parent} generation; \cite{Mazumder2016} argues that the true IGE in the US is closer to 0.6. Column (2) reports estimates based on annual incomes for the child generation, pooling observations between age 22 and the upper age indicated in each row. As in the Swedish sample, the estimates are lowest at early age, and remain substantially below the benchmark for all considered age ranges.

In columns (3) to (7) we report  the various lifecycle estimators. We keep the discussion brief as the patterns are similar as in the Swedish data (although point estimates are noisier due to the smaller samples).  Column (3) reports estimates from equation (\ref{eqtwostep}) without parental characteristics $P_{ic}$, distinguishing three education groups. As in the Swedish data, this baseline estimator understates the IGE, in particular when incomes are observed at early age. Columns (4) and (5) report the parental lifecycle estimator as defined in equation (\ref{eqtwostep}), with either linear or quadratic age interactions with parent income. These variants perform better than the baseline estimator, in particular at early age. However, they still understate the IGE if incomes are measured at a very young age, for the same reasons as illustrated in the Swedish data. Column (6) reports estimates from the parental lifecycle estimator without individual fixed effects. As for Sweden, this estimator is stable over age and close to the benchmark. Finally, column (7) reports estimates from the slope-level estimator based on equation (\ref{eqtwostepFE}). It performs better than the baseline estimator, but still varies with the age at which child incomes are measured.

Overall, the lifecycle estimator interacting a quadratic in child age with parental income, with or without fixed effects (columns 5 and 6), performs well in both the Swedish and US data. It nearly eliminates lifecycle bias in both samples, with estimates fluctuating closely around the benchmark. We showed that the mean estimates are quite stable with respect to (i) the \textit{age range} in which the child generation is observed, (ii) the \textit{number of income observations} available for each person, and (iii) the \textit{number of individuals} in the sample. This stability makes the estimator attractive for comparative purposes, such as mobility comparisons across countries or over time.

\section{Recent Trends in Income Mobility in Sweden and the US}  \label{sec:correctiontrends}

Finally, we use the lifecycle estimator to study mobility trends in Sweden and the US, with three key objectives. First, we probe whether earlier estimates may have been distorted by lifecycle effects from the use of varying age windows across cohorts (see Section \ref{sec:correctionmethods}). Second, the estimator's robustness to the age at which child incomes are observed (Section \ref{sec:newcorrection}) allows us to analyze mobility trends for younger, more recent birth cohorts not considered in previous studies, which are particularly interesting from a policy perspective. Third, this application moves beyond the ``idealized data setting'' considered earlier, making it a useful reference for other researchers interested in applying the estimator. 

\subsection{A Lifecycle Estimator for Mobility Trends} 
Because recent cohorts can only be observed at a young age, their income profiles need to be extrapolated over unobserved ages. For example, those born in 1989 are only observed up to age 29 in the Swedish data and 27 in the PSID. One way to address this issue is to pool individuals from different cohorts (as in Table A.6) and assume that, conditional on education or other observables, the shape of age-income profiles remains constant across cohorts (\citealt{Vogel2007}, \citealt{Haider2006}). However, age-income profiles may in fact change (see also \citealt{eshaghnia2022intergenerational}), and Figure A.3 shows that income growth differs not only by education but also between Swedish cohorts conditional on education. Figure A.4 provides the corresponding evidence by occupation.

To capture such changes in the shape of age-income profiles, we can extend equation (\ref{eqtwostep}) by allowing for the age interactions with own and parental characteristics $Z_{ic}$ and $P_{ic}$ to vary across cohorts. As linear interactions generally perform poorly (see Section \ref{sec:newcorrection}) and quadratic interactions might be unstable when extrapolating over wide age intervals, we instead interact decade-of-birth dummies with parental income and a ``standardized'' age profile that captures the average (concave) shape of age-income profiles in our sample. This approach allows income growth differences by parental income to vary across cohort groups but assumes that the overall shape of age-income profiles remains otherwise similar.

\subsection{Mobility Trends in Sweden} 

\singlespacing
\begin{table}[ht!]
  \centering
   \caption{Trends in Income Mobility in Sweden (Register data) }
   \begin{adjustbox}{width=\textwidth}
    \begin{tabular}{p{0.11\textwidth}>{\centering}p{0.11\textwidth}>{\centering}p{0.11\textwidth}>{\centering}p{0.0\textwidth}>{\centering}p{0.11\textwidth}>{\centering}p{0.11\textwidth}>{\centering}p{0.11\textwidth}>{\centering\arraybackslash}p{0.11\textwidth}}
    \toprule
    \toprule
          & \multicolumn{2}{c}{Direct Estimator} &       & \multicolumn{3}{c}{Lifecycle estimator} &  \\
\cmidrule{2-3}\cmidrule{5-8}          & Annual & Annual &       & Baseline  & Parental  & Parental & Parental \\
          & All ages & Age 25-30 & & FE & FE & FE & no FE \\
          &  &  &       &   --    &   --    & Cohort  & Cohort  \\
          &       &       &       &       &       & interaction & interaction \\
          & (1)   & (2)   &       & (3)   & (4)   & (5)   & (6) \\
\cmidrule{2-8}    Cohort  & 0.230 & 0.087 &       & 0.197 & 0.197 & 0.195 & 0.180 \\
        1950-59  & (0.002) & (0.002) &       & (0.002) & (0.002) & (0.002) & (0.001) \\
          &       &       &       &       &       &       &  \\
    Cohort  & 0.224 & 0.138 &       & 0.207 & 0.212 & 0.213 & 0.202 \\
      1960-69    & (0.002) & (0.002) &       & (0.002) & (0.002) & (0.002) & (0.000) \\
          &       &       &       &       &       &       &  \\
    Cohort  & 0.198 & 0.162 &       & 0.198 & 0.209 & 0.199 & 0.187 \\
     1970-79     & (0.002) & (0.002) &       & (0.002) & (0.002) & (0.002) & (0.000) \\
          &       &       &       &       &       &       &  \\
    Cohort  & 0.162 & 0.153 &       & 0.181 & 0.197 & 0.164 & 0.152 \\
      1980-89    & (0.002) & (0.002) &       & (0.002) & (0.002) & (0.002) & (0.000) \\
          &       &       &       &       &       &       &  \\
    $R^2$    & 0.025 & 0.017 &       & 0.057 & 0.061 & 0.058 & 0.326 \\
    Obs   & 39,148,343 & 9,921,334 &       & 1,844,829 & 1,844,829 & 1,844,829 & 1,844,829 \\
    Individuals & 1,844,829 & 1,842,203 &       & 1,844,829 & 1,844,829 & 1,844,829 & 1,844,829 \\
    \bottomrule
    \bottomrule
    \end{tabular}%
   \end{adjustbox}
  \label{tab:trendssweden}%
  \smallskip
   \fnote{\textbf{Notes:} Columns (1) and (2) are based on direct regressions in which we regress son's log annual income on father's lifetime income, pooling all available income observations at age 25-58 (column 1) or in a fixed age range 25-30 (column 2). Columns (3) to (6) report different variants of the lifecycle estimator based on all available income observations. Column (3) includes individual FEs and a quartic in age interacted with dummies for four education groups. Column (4) adds a quadratic interaction between son's age and fathers' income and a linear interaction between son's age and father's education. We next add interactions between parental income x cohort dummies and a standardized profile in column (5) (see main text). Finally, column (6) follows the specification of column (5), but intercepts are a function of parental income rather than individual-specific (no fixed effects).} 
\end{table}%
\doublespacing

Existing evidence on Swedish mobility trends is scarce, especially for cohorts born after the 1970s.\footnote{\cite{engzell2021robust} analyze how Swedish mobility trends vary across a number of specification choices. \cite{branden2022mother} show trends over calendar years with a focus on gender differences.} Table \ref{tab:trendssweden} reports our estimates, distinguishing four groups of cohorts born 1950-1989.\footnote{The estimated IGE for the 1950s cohorts is slightly lower than in our benchmark sample because of differences in how the samples were constructed. To keep quality constant across cohorts, parental income is measured as a shorter average in the trends sample (see Section \ref{sec:data}), introducing attenuation bias. Moreover, our benchmark sample is restricted to fathers who were relatively young at the birth of the son, for whom parental income is better observed.} The first two columns report direct estimates based on annual incomes. In column (1), we pool all available income observations in a regression of (log) annual income of children on the father's log income. As expected, these estimates \emph{decrease} monotonically across cohorts, from 0.23 for those born in the 1950s to 0.16 for cohorts born in the 1980s.  In column (2), we instead consider incomes at a fixed age range available for all cohorts, age 25-30. These estimates \emph{increase} across cohorts, by nearly 80 percent. Neither specification seems plausible. The former estimates are based on different age windows over cohorts, but estimates of the IGE tend to increase with the age at which child incomes are observed – explaining why they are larger for earlier cohorts. The latter estimates promise to address this issue by holding the age window fixed, but lifecycle profiles may differ across cohorts because of changes in educational attainment (\citealt{heckman2021lessons}) or because the education-specific profiles differ (see Figure A.3). 

Columns (3)-(6) report variants of the lifecycle estimator in equation (\ref{eqtwostep}). Column (3) presents the baseline estimator that allows age-income profiles to vary by education but not across cohorts (except for shifts in the individual intercepts $\alpha_i$ and in education). The estimates are stable over earlier cohorts but decline slightly -- by about 10 percent -- for the 1980-89 cohorts. In column (4), we add interactions between parent (log) income or education and a quadratic of child age to account for variation in income growth by parental background within education groups. This addition has little effect on IGE estimates for earlier cohorts, but increases the estimates for the more recent cohorts. As is intuitive, accounting for the steeper income growth among children from more affluent parents (see Table \ref{tab:tabsec3}) is particularly consequential for recent cohorts, for which only early-age incomes are observed. Finally, we allow age-income profiles to vary across cohorts, conditional on the other regressors in the model. In column (5) we interact log parental income with cohort-group dummies and the ``standardized'' age profile described above. The specification in column (6) is similar but drops the individual fixed effects $\alpha_i$, allowing intercepts to vary only with parental income. These variants suggest that the IGE remained stable between the 1950 and 1970s cohorts before declining for the most recent cohorts.

In sum, while estimates based on a fixed age window suggest that mobility \textit{decreased} substantially after the 1950s cohorts, accounting for lifecycle effects yields estimates that vary much less across cohorts. Indeed, our preferred estimator in column (5) suggests that mobility remained stable for cohorts born between the 1950s and 1970s, but then \textit{increased} for more recent cohorts. 
To further illustrate the divergence between the simpler and our preferred lifecycle estimator, Appendix Table F.1 shows sequentially how  estimates evolve once different predictors are added. 

These results illustrate that for the estimation of mobility trends up to more recent cohorts, it is critical to allow for income profiles to vary by education and parental income. It can also be important to allow for such relationships to vary over cohorts, as in our application. Part of the divergence between our preferred and more basic estimators can be explained by changes in the relationship between education and income growth, but changes in the relationship between parental income and own income growth within education groups also contribute, as documented in Table F.3.

\subsection{Mobility Trends in the United States} 

Many previous studies find that income mobility in the US has remained relatively stable for cohorts born between the 1950s and late 1970s (\citealt{Hertz2007}, \citealt{Lee2009}, \citealt{Chetty2014a}), though others report a decline in mobility (\citealt{Justman2021}; \citealt{jacome2021mobility}) or that trends vary across the income distribution (\citealt{Palominoetal2017}). The stability observed in many studies is puzzling, given the concurrent rise in income inequality (\citealt{Katz1999}) and the negative relation between inequality and mobility predicted by theory \citep{Solon2014} and observed across countries (\citealt{Blanden2013}; \citealt{Corak2013}) and regions within countries (\citealt{Chetty2014}; \citealt{NyStu2021Geography}). Some studies discuss why a decline in mobility has not been observed.\footnote{In particular, \cite{Davis2019} show that studies based on the PSID miss a reduction in mobility that occurred already for cohorts born in the early 1950s, who entered the labor market when inequality was rising during the 1980s. Moreover, \cite{Nybom2016a} argue that changes in the joint distribution of income and education in the \emph{parent} generation may have counteracted the effect of rising income inequality on more recent cohorts.} Others argue that it is yet to happen. For example, \cite{putnam2012growing} note that the ``\textit{adolescents of the 1990s and 2000s are yet to show up in standard studies of intergenerational mobility but [other evidence suggests] that mobility is poised to plunge dramatically.}'' We can study these cohorts, as our lifecycle estimator performs well when incomes are observed at young ages.

\begin{table}[t!]
  \centering
   \caption{Trends in Income Mobility in the US (PSID)}
  \begin{adjustbox}{width=\textwidth}
     \begin{tabular}{p{0.11\textwidth}>{\centering}p{0.11\textwidth}>{\centering}p{0.11\textwidth}>{\centering}p{0.0\textwidth}>{\centering}p{0.11\textwidth}>{\centering}p{0.11\textwidth}>{\centering}p{0.11\textwidth}>{\centering\arraybackslash}p{0.11\textwidth}}
    \toprule
    \toprule
          & \multicolumn{2}{c}{Direct Estimator} &       & \multicolumn{3}{c}{Lifecycle estimator} &  \\
\cmidrule{2-3}\cmidrule{5-8}          & Annual & Annual &       & Baseline  & Parental  & Parental & Parental \\
          & All ages & Age 25-30 & & FE & FE & FE & no FE \\
          &  &  &       &   --    &   --    & Cohort  & Cohort  \\
          &       &       &       &       &       & interaction & interaction \\
          & (1)   & (2)   &       & (3)   & (4)   & (5)   & (6) \\
\cmidrule{2-8}    Cohorts & 0.380 & 0.309 &       & 0.417 & 0.430 & 0.437 & 0.434 \\
        1950-59  & (0.034) & (0.039) &       & (0.040) & (0.040) & (0.040) & (0.017) \\
        \noalign{\medskip}
    Cohorts & 0.391 & 0.361 &       & 0.429 & 0.447 & 0.438 & 0.449 \\
        1960-69   & (0.034) & (0.037) &       & (0.036) & (0.035) & (0.036) & (0.013) \\
        \noalign{\medskip}
    Cohorts & 0.406 & 0.349 &       & 0.450 & 0.468 & 0.470 & 0.475 \\
        1970-79   & (0.029) & (0.033) &       & (0.033) & (0.033) & (0.033) & (0.009) \\
        \noalign{\medskip}
    Cohorts  & 0.308 & 0.311 &       & 0.363 & 0.394 & 0.432 & 0.443 \\
        1980-89  & (0.025) & (0.027) &       & (0.025) & (0.025) & (0.026) & (0.008) \\
        \noalign{\medskip}
     $R^2$     & 0.087 & 0.082 &       & 0.147 & 0.159 & 0.165 & 0.581 \\
    Obs   & 59,458 & 17,616 &       & 4,937 & 4,937 & 4,937 & 4,938 \\
    Individuals & 4,939 & 4,565 &       & 4,937 & 4,937 & 4,937 & 4,938 \\
     \bottomrule
    \bottomrule
    \end{tabular}%
 \end{adjustbox}
  \label{tab:trendspsid}%
  \smallskip
 \fnote{\textbf{Notes:} Columns (1) and (2) are based on direct regressions in which we regress offspring's log annual income on parental lifetime income. For column (1) we pool all available income observations at age 22-58. In column (2) we only consider age 25-30. Columns (3) to (6) report different variants of the lifecycle estimator based on all available income observations. Column (3) includes individual FEs and a quartic in age interacted with dummies for three education groups. Column (4) adds a quadratic interaction between son's age and fathers' income and between son's age and father's education. We next add interactions between parental income x cohort dummies and a standardized profile in column (5) (see main text). Finally, column (6) follows the specification of column (5), but intercepts are a function of parental income rather than individual-specific (no fixed effects).}
\end{table}%

Specifically, we use the PSID to estimate the IGE for four different cohort groups born in the 1950s, 1960s, 1970s or 1980s, using the trends sample described in Section \ref{sec:data}. Table \ref{tab:trendspsid} reports the results, following the same structure as Table \ref{tab:trendssweden}. The first two columns present ``naive'' regressions in which we regress log annual income on log parental income. Pooling all observed incomes (column 1), yields lower IGE estimates for more recent cohorts, which are observed only at a young age (generating a downward bias). Holding the age window fixed (column 2) suggests that mobility decreased for the 1960s and 70s cohorts, but rebounded thereafter. However, as already discussed, neither of these estimators is reliable. Switching to a lifecycle estimator generally yields larger estimates (columns 3-6). In column (3), the baseline estimator that does not account for differential income growth by parental background indicates a slight increase in the IGE between the 1950s and 1970s cohorts (in line with \citealt{Justman2021}), but a sizable drop for the 1980s cohorts. 

Allowing income growth to vary with parental income or education (column 4) has little effect on earlier cohorts, but increases the IGE estimate for the 1980s cohorts. This pattern is consistent with the evidence shown in Section \ref{sec:incprocess}: allowing for differential lifecycle growth is particularly important if individuals are only observed at young ages. Moreover, the shape of income trajectories might have changed across cohorts. In column (5) we account for such shifts by interacting indicators of the four cohort groups with parental income and a ``standardized'' age profile (as defined above).  This increases the estimated IGE for recent cohorts further, from 0.394 to 0.432. Finally, column (6) shows that these results do not depend on the inclusion of individual fixed effects. 

In sum, all variants of the lifecycle estimator suggest that mobility decreased only slightly between the 1950s and 1970s cohorts, but the pattern for more recent cohorts depends on whether we account for changes in income \textit{growth} across cohorts. Naive estimators suggest that mobility increased markedly for cohorts born in the 1980s, while estimators that allow for differential lifecycle growth suggest no significant change in mobility. 

As for Sweden, we systematically compare the role of specific income-growth predictors in Table F.2. Compared to our preferred estimate in the final column, the estimate for the 1980s cohorts is strongly underestimated when not allowing for heterogeneity in income growth. It remains underestimated, though less so, once we introduce cohort-constant heterogeneity with respect to parental income (col. 3) or cohort-varying heterogeneity with respect to education (col. 4 and 5). But it is only when introducing cohort-varying heterogeneity with respect to parental income (col. 6) that we find a pattern of stable mobility between the 1950s and 1980s cohorts. Similar to the Swedish case, the estimated trends depend thus both on changes in the relationship between own education and future income, and on changes in the relationship between own income growth and parental income conditional on education. 

Structural changes on the labor market may have therefore not only affected the distribution of income at a given age, but also the distribution of income \textit{growth}: in the US, children from richer parents experience faster income growth over age today than in the past. Of course, these findings are only a snapshot based on early labor market experiences, and it remains to be seen whether they hold up when the 1980s cohorts reach later stages of their careers. Based on current data, however, we reject any major change in US income mobility over the past four decades.

\section{Concluding Remarks}  \label{sec:conclusion}

Intergenerational mobility in income is difficult to measure, and methodological improvements have led to major revisions of mobility estimates (\citealt{Solon1999,Mazumder2016}). But despite a better understanding of the source of attenuation and lifecycle biases, the literature still struggles to address them effectively. As \cite{mogstad2021} note, ``\textit{there is considerable uncertainty associated with the IGE estimates, and especially with their comparison across time and place}''. 
The commonly used rule of thumb to measure income around midlife only helps partially, as it cannot be applied to recent cohorts, and estimates remain sensitive to the exact age at measurement.

Instead, we proposed that researchers make more systematic use of available income information over the lifecycle. We highlighted three properties of income processes relevant for intergenerational research: (i) income growth explained by observable characteristics, (ii) transitory noise, and (iii) unexplained income growth that correlates within families. The latter is also of more general interest for a long-standing debate on whether (residual) income grows at an individual-specific and deterministic rate or follows a random walk. Using long income series from Sweden and the US, we found that residual income growth contains a systematic component: children from affluent parents tend to experience faster growth, even accounting for their own characteristics. 

The estimation of intergenerational mobility is therefore intertwined with the analysis of income dynamics. Building on earlier contributions such as \cite{Hertz2007}, we proposed a two-step \textit{lifecycle} estimator of income mobility that captures these dynamics. In the first step, we predict income profiles based on individual characteristics, but allow income growth to also vary with parental background. Comparing this lifecycle estimator to benchmark estimates in Swedish and US data, we found it to perform well in different data settings, and to be less sensitive to the age at which incomes are observed than other methods. These properties are attractive for comparative purposes, such as mobility comparisons across place or time, or for studying mobility in recent birth cohorts that are observed only at younger ages. Our main analysis employs a parametric first-step model, which outperforms more data-driven methods in our setting. Future research could explore the potential of more tailored implementations of machine learning approaches.

We used this estimator to study mobility trends in Sweden and the US, including for more recent birth cohorts, which are particularly interesting from a policy perspective. For Sweden, estimates based on a fixed age window suggest that mobility decreased substantially after the 1950s cohorts. However, after accounting for lifecycle effects, we find that the IGE has remained stable for cohorts born in the 1950s, 1960s and 1970s, and decreased slightly for more recent cohorts. 

Accounting for lifecycle effects is particularly important for recent US cohorts. While a naive fixed-age estimator yields a U-shaped mobility pattern across cohorts, our lifecycle estimator yields more stable estimates: the IGE has been remarkably constant across US cohorts born 1950-1989. Interestingly, income growth has diverged in more recent cohorts, with children from more affluent backgrounds experiencing faster income growth today than in the past. One explanation for this finding is an increased divergence in income growth within education groups. An interesting question for future work is \textit{why} children from affluent families experience faster income growth, even conditional on education and other characteristics, and why these patterns change over time. 

\singlespacing
\bibliographystyle{aer}
\bibliography{refMNS.bib}

\clearpage
\newpage
\appendix
\onehalfspacing

\newpage
\begin{center}
\section*{\Huge Online Appendix} 
{\large \textbf{\thetitle \\
\vspace{1em}
 \theauthor}}
\end{center}
\setcounter{page}{1}

\section{Additional Figures and Tables}\label{app:fig_tab}
\setcounter{table}{0}
\renewcommand{\thetable}{A.\arabic{table}}
\setcounter{figure}{0}
\renewcommand{\thefigure}{A.\arabic{figure}}

\begin{figure}[h!]
\caption{\label{fig:figA1} Components of the Income Process}
\centering
\includegraphics[width=1\textwidth]{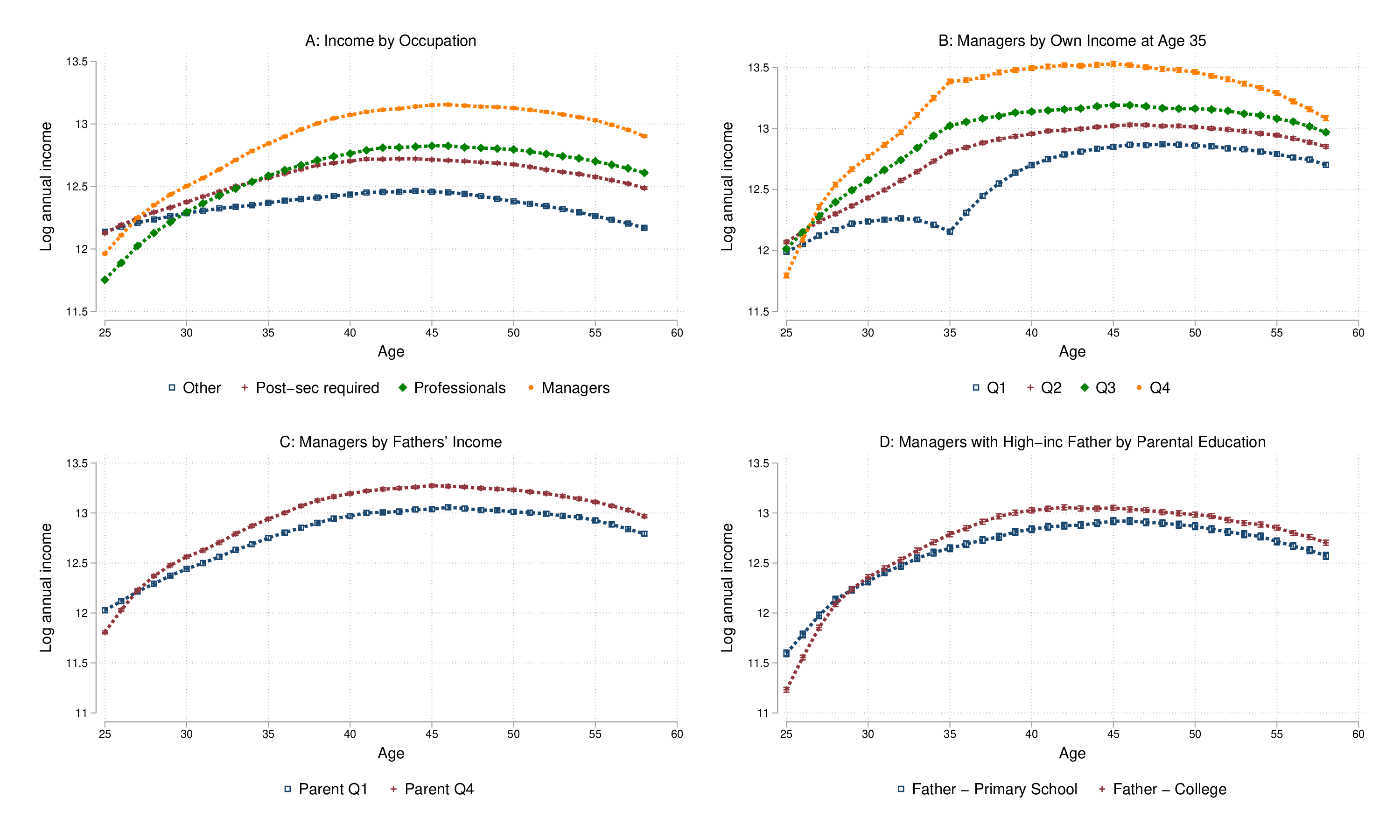}
\fnote{\textbf{Notes:} Panel A shows income trajectories by type of occupation. Panel B focuses only on managers, who are split into four groups, according to their annual income at age 35. Category Q1 refers to the bottom and Q4 to the top quartile. In Panel C, managers are divided into two groups, those at the top quartile and those at the bottom quartile of fathers' lifetime income. Finally, in Panel D, managers whose fathers belong to the top half of lifetime income are divided into two additional groups: college-educated fathers and fathers with only primary school. We remove time effects from annual income observations to abstract from the business cycle. Confidence intervals (95\%) are plotted around each line.}
\end{figure}

\begin{figure}[!ht]
\caption{\label{fig:figA2} Income Profiles by Own Lifetime Income}
\centering
\includegraphics[width=1\textwidth]{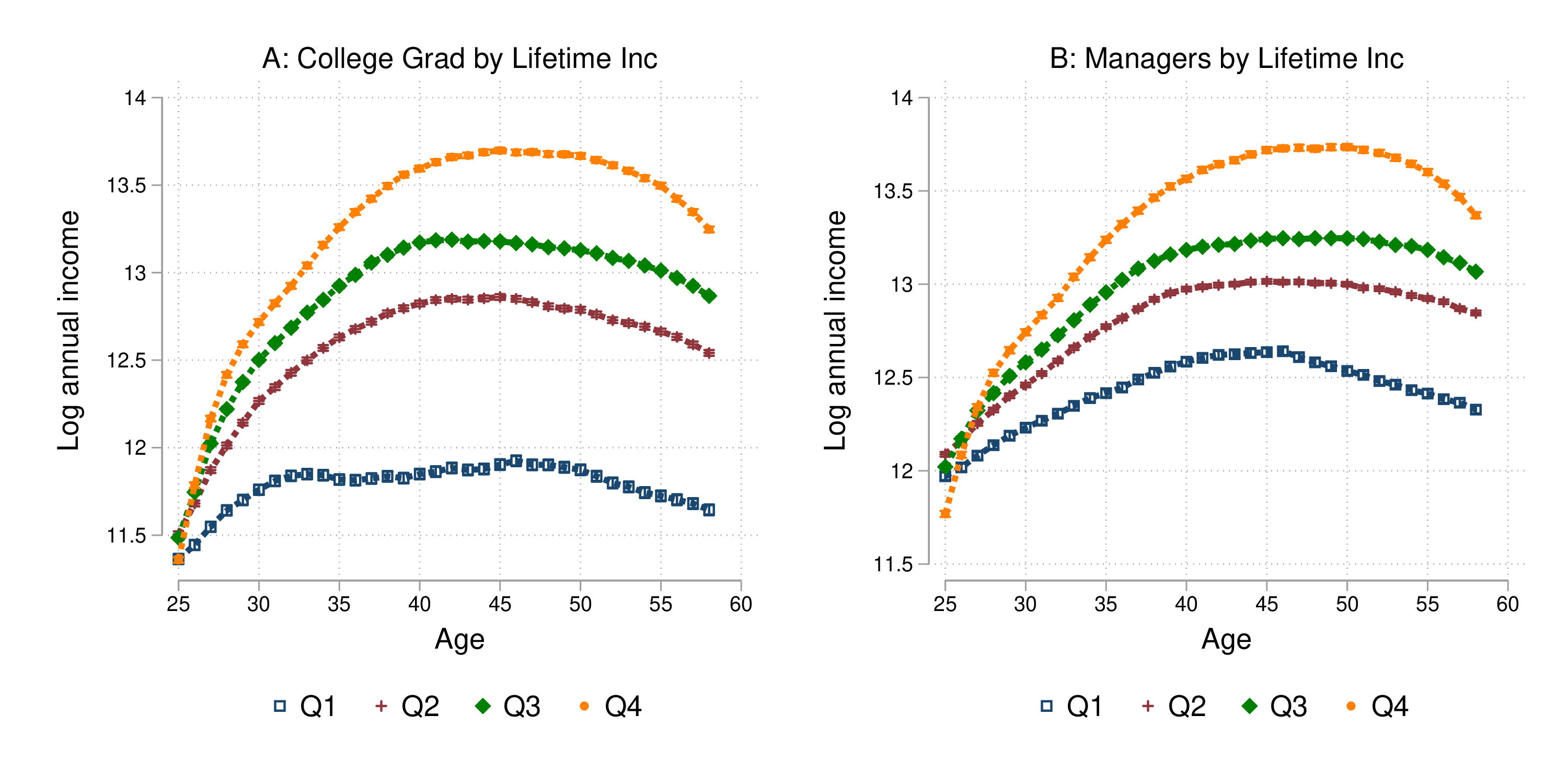}
\fnote{\textbf{Notes:} 
Panel A (Panel B) shows the income trajectories of college graduates (managers) split into four quartiles of their own lifetime income. We remove time effects to abstract from the business cycle. Confidence intervals (95\%) are plotted around each line.}
\end{figure} 

\begin{figure}[h!]
 \centering
 \caption{\label{fig:educ_coh}{Income Profiles by Education Group and Cohort}}
   \includegraphics[width=1\textwidth]{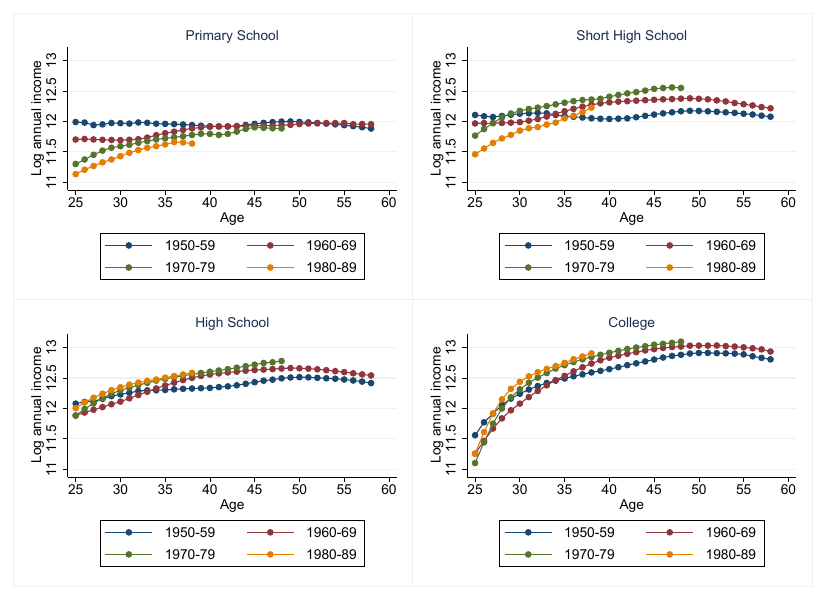}
    \fnote{\textbf{Notes:} The figure plots the observed log income profiles by education group and cohort. Source: Swedish Register Data.}
\end{figure}

\begin{figure}[h!]
 \centering
 \caption{\label{fig:occ_cohort}{Income Profiles by Occupation Group and Cohort}}
   \includegraphics[width=1\textwidth]{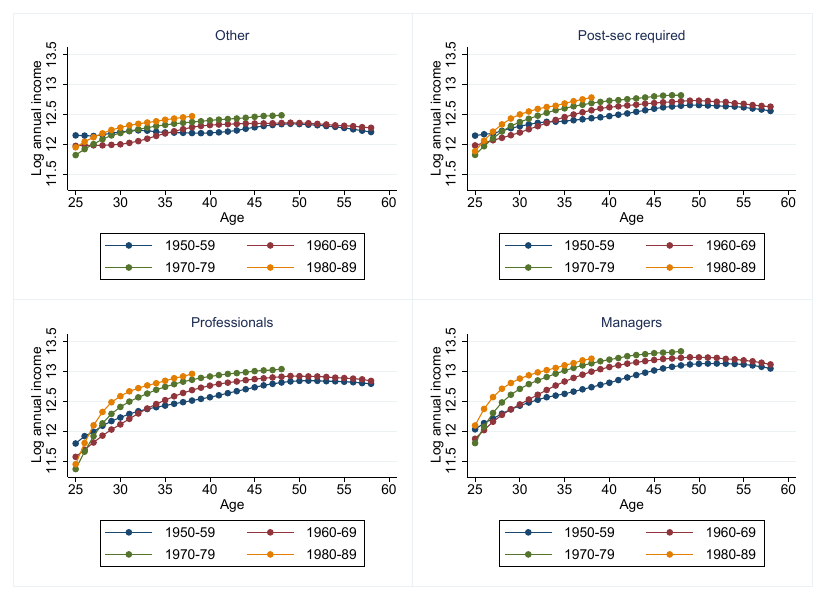}
    \fnote{\textbf{Notes:} The figure plots the observed log income profiles by occupation group and cohort. Source: Swedish Register Data.}
\end{figure}

\clearpage

\begin{sidewaystable}
  \centering
  \caption{Intergenerational Elasticity Literature}
   \begin{adjustbox}{width=\textwidth}
    \begin{tabular}{p{14.665em}llll}
    \toprule
        \toprule
    \multicolumn{1}{l}{\textbf{Authors}} & \textbf{Journal } & \textbf{Estimate (US)} & \textbf{Method} & \textbf{Addressing lifecycle bias in offspring generation} \\
    \midrule
    \multicolumn{1}{l}{\textbf{IGE in Levels}} &       &       &       &  \\
    \midrule
    \cite{Solon1992}  & AER   & 0.41  & Averaging & \multicolumn{1}{p{25.75em}}{Single year annual earnings, average age 29.6} \\
    \cite{Zimmerman1992}  & AER   & 0.54  & Averaging & \multicolumn{1}{p{25.75em}}{Single year of son's annual earnings; average age 33.8 } \\
    \cite{Mazumder2005}  & Restat & 0.61  & Averaging & Average 4 years of income, age 30-35 \\
    \cite{Hertz2006}   & Working Paper & 0.58  & Averaging & Average 4.1 income observations at mean-age 37 \\
    \cite{Bratsberg2007}  & Economic Journal & 0.54  & Averaging & Annual income in 1995 and 2001, cohorts 1957-1964 \\
    \cite{Gouskova2010}  & Labour Economics & 0.63  & Averaging & Single year from ages 35-44 \\
    \cite{Chau2012}  & Economic Letters & 0.66  & Model Income & \multicolumn{1}{p{25.75em}}{At least 3 observations of annual earnings between the ages of 25-60. Use earnings dynamics model} \\
    \cite{Jantti2012}  & Economic Letters & Sweden & Model Income & \multicolumn{1}{p{25.75em}}{Formulate simple model with heterogeneous income } \\
    \cite{Chetty2014}  & QJE   & 0.34  & Averaging & 2-year average around ages 29-32 (2011 and 2012) \\
    \cite{Mitnik2015}  & Working Paper & 0.56  & Averaging & Single year around ages 35-38 \\
    \cite{Mazumder2016}  & Research in Labor Economics & 0.66  & Averaging & Average between 1 and 11 years around age 40 \\
    \cite{Borisov2016}  & Working Paper & Russia & Model Income & \multicolumn{1}{p{25.75em}}{Predicted value of permanent earnings based on monthly earnings. Controls for hours worked, age, year of birth, education} \\
    \cite{Landers2017}  & Scandinavian Journal & 0.29 to 0.45 & Averaging & \multicolumn{1}{p{25.75em}}{Average between ages 34-41 for older cohorts down to 30-35 for younger cohorts} \\
    \cite{Deutscher2020}  & Labour Economics & Australia & Averaging & \multicolumn{1}{p{25.75em}}{Average over five years around ages 29-37 }  \\ 
     \cite{Connolly2021}  & NBER Book Chapter & Canada & Averaging & \multicolumn{1}{p{25.75em}}{Average over ages 30 to 36 } \\
    \midrule
    \textbf{IGE in Trends }     &       &       &       &  \\
    \midrule
    \cite{Mayer2005}  & Journal of Human Resources & Non-linear & Averaging & Son's family income at age 30 \\
    \cite{Hertz2007} & Industrial Relations & No trend & Model Income & Estimation of income profiles \\
    \cite{Aaronson2008}  & Journal of Human Resources & Non-linear & --     &  \\
    \cite{Lee2009}  & Restat & No trend & Averaging & Average of all available years, changing across cohorts \\
    \cite{Justman2013} & Working Paper & Upward & \multicolumn{1}{c}{Model Income} & \multicolumn{1}{l}{Predicted income at age 40, controls for age, education, race, marital status and individual FE} \\
   \cite{Hartley2017} & JPE & Upward & Averaging & \multicolumn{1}{p{25.75em}}{Multi-year average, Lee \& Solon age adjustment (mother-daughters)} \\
   \cite{Davis2019}  & Restat & Upward & Averaging & \multicolumn{1}{p{25.75em}}{3-year average of total family income when sons were 29 and spouses 28 on average} \\
    \midrule
     \textbf{Lifecycle Bias (Methodological)}     &       &       &       &  \\
    \midrule
    \cite{Haider2006}  & AER   & --     & --     & \multicolumn{1}{p{25.75em}}{Generalized errors-in-variables (GEiV) model} \\
    \cite{Grawe2006} & Labour Economics & --     & --     & \multicolumn{1}{p{25.75em}}{Discussion of lifecycle bias} \\
   \cite{Bohlmark2006}  & Journal of Labor Economics & --     & --     & \multicolumn{1}{p{25.75em}}{Discussion of lifecycle biase and GEiV model} \\
    \cite{Nilsen2012}  & Scandinavian Journal & --     & --     & \multicolumn{1}{p{25.75em}}{Discussion of lifecycle bias} \\
   \cite{Nybom2016}  & Journal of Human Resources & --     & --     &  \multicolumn{1}{p{25.75em}}{Testing lifecycle bias and GeiV model} \\
   \cite{Chen2017} & Labour Economics & --     & --     & Discussion of lifecycle bias \\
    \cite{Gregg2017} & Oxford Bulletin & --     & --     & Discussion of lifecycle bias and application for UK \\
    \bottomrule
    \bottomrule
    \end{tabular}%
       \end{adjustbox}
  \label{tab:literature}%
  \fnote{\textbf{Notes:} This table presents recent papers that attempt to measure the intergenerational elasticity in levels or in trends with a brief description of the main approach used to address the lifecycle bias in the measurement of offspring income. It also contains some methodological papers that discuss and test the lifecycle bias in intergenerational mobility estimates.}
\end{sidewaystable}

\begin{table}[h!]
  \centering
\caption{Associating income growth of sons with income growth and levels of fathers}
    \begin{tabular}{lrrcc}
    \toprule
    \toprule
          & \multicolumn{1}{c}{(1)} & \multicolumn{1}{c}{(2)} & (3)   & (4) \\
    \midrule
    \multicolumn{5}{c}{Dependent variable: Change in son's log annual income between ages 25 and 30} \\
    \midrule
    Change in father's log annual  & \multicolumn{1}{c}{0.096***} & \multicolumn{1}{c}{0.082***} & 0.027*** & 0.014*** \\
    income between ages 25 and 30 & \multicolumn{1}{c}{(0.003)} & \multicolumn{1}{c}{(0.004)} & (0.003) & (0.003) \\
    Log (Father's Lifetime  & \multicolumn{1}{c}{} & \multicolumn{1}{c}{9.454***} & 1.994*** & 0.651 \\
    Income)/100 & \multicolumn{1}{c}{} & \multicolumn{1}{c}{(0.398)} & (0.391) & (0.398) \\
    \midrule
    N     & \multicolumn{1}{c}{170,044} & \multicolumn{1}{c}{170,044} & 170,044 & 170,044 \\
    $R^2$  & \multicolumn{1}{c}{0.005} & \multicolumn{1}{c}{0.009} & 0.100 & 0.102 \\
    \midrule
    \midrule
    \multicolumn{5}{c}{Dependent variable: Change in son's log annual income between ages 30 and 35} \\
    \midrule
    Change in father's log annual  & \multicolumn{1}{c}{0.008} & \multicolumn{1}{c}{0.007} & 0.003 & 0.002 \\
    income between ages 30 and 35 & \multicolumn{1}{c}{(0.004)} & \multicolumn{1}{c}{(0.004)} & (0.004) & (0.004) \\
    Log (Father's Lifetime  & \multicolumn{1}{c}{} & \multicolumn{1}{c}{0.475} & -1.409** & -1.598*** \\
    Income)/100 & \multicolumn{1}{c}{} & \multicolumn{1}{c}{(0.422)} & (0.432) & (0.442) \\
    \midrule
    N     & \multicolumn{1}{c}{104,393} & \multicolumn{1}{c}{104,393} & 104,393 & 104,393 \\
    $R^2$  & \multicolumn{1}{c}{0.000} & \multicolumn{1}{c}{0.000} & 0.009 & 0.009 \\
    \midrule
    \midrule
    \multicolumn{5}{c}{Dependent variable: Change in son's log annual income between ages 35 and 40} \\
    \midrule
    Change in father's log annual  & \multicolumn{1}{c}{-0.003} & \multicolumn{1}{c}{-0.005} & -0.005 & -0.005 \\
    income between ages 35 and 40 & \multicolumn{1}{c}{(0.008)} & \multicolumn{1}{c}{(0.008)} & (0.008) & (0.008) \\
    Log (Father's Lifetime  & \multicolumn{1}{c}{} & \multicolumn{1}{c}{0.964} & 0.220 & 0.375 \\
    Income)/100 & \multicolumn{1}{c}{} & \multicolumn{1}{c}{(0.867)} & (0.880) & (0.909) \\
    \midrule
    N     & \multicolumn{1}{c}{23,447} & \multicolumn{1}{c}{23,447} & 23,447 & 23,447 \\
    $R^2$  & \multicolumn{1}{c}{0.000} & \multicolumn{1}{c}{0.000} & 0.003 & 0.003 \\
    \midrule
    Education &       &       & X     & X \\
    Father's Education &       &       &       & X \\
    \bottomrule
    \bottomrule
    \end{tabular}%
 \label{tab:hipgrowth}%
   \smallskip
   \fnote{\textbf{Notes:} Birth cohorts 1970-78. The dependent variable in each panel is the change in log annual income over the indicated age range. Education distinguishes 12 levels of highest educational attainment. Father's education distinguishes four levels of parental education. \sym{*} $p<0.05$, \sym{{*}{*}} $p<0.01$, \sym{{*}{*}{*}} $p<0.001$.}
\end{table}%

\begin{table}[h!]
  \centering
\caption{IGE Estimates Accounting for Age-Education Profiles}
    \begin{tabular}{rcccc}
    \toprule
    \toprule
    \multicolumn{1}{l}{{}} & \multicolumn{3}{c}{\underline{Panel A}} & \multicolumn{1}{c}{\underline{Panel B}} \\
    \multicolumn{1}{l}{{Observed}} & \multicolumn{3}{c}{Prediction at Age} & \multicolumn{1}{c}{{Prediction of}} \\
        \multicolumn{1}{l}{{Age Range}} & \textit{25} & \textit{30} & \textit{35} &  Complete Profiles\\
    \midrule
    \multicolumn{1}{p{5em}}{25-30} & 0.042 & 0.165 & - & 0.209 \\
          & (0.003) & (0.003) &       & (0.004) \\
    \multicolumn{1}{l}{\textit{25-35}} & 0.080 & 0.158 & 0.224 & 0.231 \\
          & (0.003) & (0.003) & (0.003) & (0.004) \\
    \multicolumn{1}{l}{\textit{25-40}} & 0.121 & 0.173 & 0.224 & 0.249 \\
          & (0.003) & (0.003) & (0.003) & (0.004) \\
    \multicolumn{1}{l}{\textit{25-45}} & 0.147 & 0.190 & 0.226 & 0.259 \\
          & (0.003) & (0.003) & (0.003) & (0.005) \\
    \multicolumn{1}{l}{\textit{25-58}} & 0.182 & 0.224 & 0.248 & 0.277 \\
          & (0.003) & (0.003) & (0.003) & (0.004) \\
    \midrule
    \multicolumn{1}{l}{\textit{Annual}} & 0.001 & 0.172 & 0.253 & - \\
          & (0.003) & (0.004) & (0.004) &  \\
    \multicolumn{1}{l}{\textit{True}} & 0.253 & 0.253 & 0.253 & 0.253 \\
          & (0.003) & (0.003) & (0.003) & (0.003) \\
     \bottomrule
    \bottomrule
    \end{tabular}%
  \label{tab:hertzvogeltab}%
  \smallskip
   \fnote{\textbf{Notes:} Benchmark sample from Swedish registers, cohorts 1952-60, N =197,242 observations. The top rows report estimates of the IGE based on the first-step estimation of equation (\ref{eqvogelhertz}), which includes a quartic in age interacted with four education groups. In Panel A, we predict child income at age 25, 30 or 35 (within the observed range). In Panel B, we predict child income over the entire lifecycle (by randomly assigning each observation of the benchmark generation into a ``young'' or an ``old'' copy, as explained in Section \ref{sec:lifecyclestimator}).}
\end{table}%

\begin{table}[h!]
  \centering
 \caption{The Lifecycle Estimator with Fewer Income Observations (Swedish data)}  
    \begin{tabular}{lcccccc}
    \toprule
    \toprule
          &       & \multicolumn{5}{c}{Lifecycle estimator (Parental, Quadratic)} \\
\cmidrule{3-7}    Son's Age & N     & $\leq$  6 obs. & $\leq$  5 obs. & $\leq$ 4 obs. & $\leq$ 3 obs. & $\leq$ 2 obs. \\
    \midrule
    Age $\leq$  30 & 94,194 & 0.287 & 0.285 & 0.285 & 0.287 & 0.282 \\
          &       & (0.003) & (0.003) & (0.004) & (0.004) & (0.004) \\
    $R^2$    &       & 0.069 & 0.067 & 0.065 & 0.063 & 0.056 \\
          &       &       &       &       &       &  \\
    Age $\leq$  35 & 94,264 & 0.261 & 0.265 & 0.263 & 0.264 & 0.264 \\
          &       & (0.003) & (0.003) & (0.003) & (0.004) & (0.004) \\
    $R^2$    &       & 0.060 & 0.060 & 0.057 & 0.055 & 0.049 \\
    \midrule
    \midrule
          &       & \multicolumn{5}{c}{Lifecycle estimator (Slope-level, Quadratic)} \\
\cmidrule{3-7}    Son's Age & N     & $\leq$ 6 obs. & $\leq$ 5 obs. & $\leq$ 4 obs. & $\leq$ 3 obs. & $\leq$ 2 obs. \\
    \midrule
    Age $\leq$  30 & 94,194 & 0.241 & 0.239 & 0.237 & 0.235 & 0.230 \\
          &       & (0.005) & (0.005) & (0.005) & (0.005) & (0.005) \\
    $R^2$    &       & 0.027 & 0.026 & 0.026 & 0.025 & 0.023 \\
          &       &       &       &       &       &  \\
    Age $\leq$  35 & 94,264 & 0.245 & 0.244 & 0.242 & 0.244 & 0.246 \\
          &       & (0.004) & (0.004) & (0.004) & (0.005) & (0.005) \\
    $R^2$    &       & 0.033 & 0.032 & 0.031 & 0.030 & 0.028 \\
    \bottomrule
    \bottomrule
   \end{tabular}
    \label{tab:lifecyclefew}%
    \smallskip
 \fnote{\textbf{Notes:} The table reports the slope coefficient from a regression of son's income on father's lifetime income for the Parental Quadratic FE and the Slope-level FE lifecycle estimators. The measure for son's income is the predicted lifetime income from a lifecycle estimator applied to the indicated age range. The top row indicates the maximum number of income observations used for each person in the child generation. The observations are selected randomly from all the observations available for each person below the indicated age threshold. Robust standard errors in parentheses.}
\end{table}%

\begin{table}[htbp]
  \centering
  \caption{The Lifecycle Estimator in Smaller Samples (Swedish data)}
    \begin{tabular}{lccccc}
    \toprule
    \toprule
    Sample Size & k=1/4 & k=1/16 & k=1/64 & k=1/256 & k=1/1024 \\
    \midrule
          & \multicolumn{5}{c}{Son's Age 25-30} \\
    \midrule
    Benchmark & 0.259 & 0.258 & 0.263 & 0.257 & 0.269 \\
    \multicolumn{1}{r}{Std. Deviation} & (0.003) & (0.011) & (0.028) & (0.054) & (0.103) \\
    Lifecycle (Parental Quadratic) & 0.249 & 0.263 & 0.266 & 0.256 & 0.266 \\
    \multicolumn{1}{r}{Std. Deviation} & (0.005) & (0.024) & (0.060) & (0.121) & (0.316) \\
     Lifecycle (Slope-level Quadratic) & 0.242 & 0.254 & 0.241 & 0.256 & 0.267 \\
    \multicolumn{1}{r}{Std. Deviation} & (0.003) & (0.022) & (0.056) & (0.113) & (0.221) \\
    \midrule
    N     & 1,711,263 & 426,585 & 106,900 & 26,723 & 6,678 \\
    \midrule
    \midrule
          & \multicolumn{5}{c}{Son's Age 25-35} \\
    \midrule
    Benchmark & 0.261 & 0.258 & 0.258 & 0.257 & 0.264 \\
    \multicolumn{1}{r}{Std. Deviation} & (0.005) & (0.009) & (0.022) & (0.050) & (0.107) \\
    Lifecycle (Parental Quadratic) & 0.266 & 0.259 & 0.266 & 0.252 & 0.253 \\
    \multicolumn{1}{r}{Std. Deviation} & (0.011) & (0.022) & (0.053) & (0.098) & (0.261) \\
     Lifecycle (Slope-level Quadratic) & 0.258 & 0.252 & 0.262 & 0.255 & 0.269 \\
    \multicolumn{1}{r}{Std. Deviation} & (0.006) & (0.017) & (0.046) & (0.084) & (0.200) \\
    \midrule
    N     & 1,711,373 & 428,891 & 106,984 & 26,722 & 6,673 \\
     \bottomrule
    \bottomrule
    \end{tabular}%
   \label{tab:sample}%
   \smallskip
   \fnote{\textbf{Notes:} The table reports the slope coefficient from a regression of son's income on father's lifetime income, comparing the Parental Quadratic FE and the Slope-level FE lifecycle estimators with the benchmark. Each column reports coefficients estimated from multiple draws with replacement of differently sized sub-samples, as indicated in the top row. For each sample size, we report the mean and standard deviation (in parentheses) of the point estimates, computed across the random draws from the main sample. Thus, for $k=\{1/4, 1/16, 1/64, 1/256, 1/1024\}$ we draw $1/k$ samples of size $N_k=N*k$ from the whole sample of size $N$. Sons' incomes are observed from age 25 to 30 in Panels A and B.}
\end{table}%

\begin{table}[htbp]
  \centering
  \caption{Robustness to Cohort and Year Effects (Swedish data)}
\begin{tabular}{lcccc}
    \toprule
    \toprule
 & (1) & (2) & (3) & (4)\tabularnewline
   \midrule
Benchmark & 0.215 & 0.215 & 0.214 & 0.210\tabularnewline
& (0.002) & (0.002) & (0.002) & (0.003)\tabularnewline
Annual, all observed ages & 0.237 & 0.237 & 0.280 & 0.288\tabularnewline
 & (0.002) & (0.002) & (0.003) & (0.006)\tabularnewline
 Lifecycle (Parental Quadratic) & 0.208 & 0.203 & 0.235 & 0.286\tabularnewline
& (0.003) & (0.003) & (0.003) & (0.005)\tabularnewline
Lifecycle (slope-level Quadratic) & 0.201 & 0.198 & 0.204 & 0.245 \tabularnewline
 & (0.002) & (0.002) & (0.003) & (0.005)\tabularnewline
   \midrule
    \textit{First-step sample} & & & & \tabularnewline
  \hspace{0.3cm}  Cohorts: & 1952-1960 & 1950-1989 & 1950-1989 & 1950-1989
  \tabularnewline
   \hspace{0.3cm}  Income years: & 1977-2018 & 1977-2018 & 1998-2018 & 2014-2018    \tabularnewline
 \midrule
Individuals (second step) & 293,333 & 293,333 & 284,806 & 152,412\tabularnewline
    \bottomrule
    \bottomrule
      \end{tabular}%
   \label{tab:app_cohort_effects}%
   \smallskip
   \fnote{\textbf{Notes:} The table reports the slope coefficient from a regression of son's income on father's lifetime income, comparing the Parental Quadratic FE and the Slope-level FE lifecycle estimators with the benchmark and an estimate based on all observed annual earnings. Each column reports estimates for cohorts born 1952-1960 using data from different time periods and (for the first-step estimation) different cohorts. We use the Swedish trends sample (see Section 2), which results in slightly lower benchmark estimates than for the baseline sample. Column (1) uses the benchmark cohorts (born 1952-1960) and all earnings years (when aged 25-58) in steps 1 and 2. Column (2) uses all cohorts (1950-1989) and all earnings years (when aged 25-58) in step 1. Column (3) uses all cohorts (when aged 25-58) during the years 1998-2018 in step 1. Column (4) uses all cohorts (when aged 25-58) during the years 2014-2018 in step 1. Robust standard errors are in parentheses and the final row shows the number of unique individuals used in each column (among the 1952-1960 cohorts).}
\end{table}%

\clearpage

\setcounter{table}{0}
\renewcommand{\thetable}{B.\arabic{table}}
\setcounter{figure}{0}
\renewcommand{\thefigure}{B.\arabic{figure}}

\section{Modelling Errors-in-Variables} \label{app:errorsinvar}

In the classical errors-in-variables model, inconsistencies in the IGE are limited to attenuation bias caused by the imprecise measurement of the lifetime income of parents (e.g., \citealt{Atkinson1980}).\footnote{While this bias can be reduced by averaging over longer income snapshots, \cite{Mazumder2005} demonstrates that even 10-year averages are not sufficient because the transitory component of income is highly serially correlated.} However, the association between current and lifetime income varies systematically over the life cycle, contrary to a classical errors-in-variables model in which the errors are independent of true values. As a consequence, the use of short income snapshots for the child generation introduces a \textit{lifecycle bias} in mobility estimates \citep{Jenkins1987}. \cite{Grawe2006} and \cite{Haider2006} demonstrate that this bias tends to be large, such that mobility estimates are quite sensitive to the age at which child income is being measured.\footnote{This observation also motivates the recent interest in mobility in income \textit{ranks}, as rank correlations suffer less from attenuation and lifecycle bias (\citealt{Chetty2014}; \citealt{Nybom2017}).}

Recent applications adopt therefore a \textit{generalized errors-in-variables} (GEiV) model proposed by \cite{Haider2006}, which accounts for the systematic relation between annual and lifetime income over the lifecycle.\footnote{The GEiV model has been extended in subsequent work. \cite{Lee2009} adapt it for the study of mobility trends. \cite{An2017} implement it within a non-parametric framework that allows for the IGE to be heterogeneous.} 
Focusing on left-hand side measurement error, it corresponds to the linear projection
\begin{equation} \label{eq:eq1}
y_{sit}=\lambda_{st}y^*_{si}+u_{sit},
\end{equation} 
where $y_{sit}$ is the annual log income of the child of family \textit{i} at age \textit{t}, $y^*_{si}$ is his or her log lifetime income, and $y^*_{si}$ and $u_{sit}$ are uncorrelated by construction. Under the assumption that $Cov(y^*_{fi}, u_{sit})=0$, with $y^*_{fi}$ denoting parental log lifetime income, the probability limit of a regression of $y_{sit}$ on $y^*_{fi}$ is 
\begin{equation} \label{eq:eq3}
plim \: \hat{\beta_t}=\frac{Cov(y_{sit}, y^*_{fi})}{Var(y^*_{fi})}=\beta\lambda_{st}, 
\end{equation}
where $\beta$ is the true IGE from regressing $y^*_{si}$ on $y^*_{fi}$. The use of short income spans would therefore not introduce bias if child income were measured at an age at which $\lambda_{st}$ is close to one, which tends to be around midlife.\footnote{\citealt{Bohlmark2006} confirm this prediction in Swedish data. As noted by \cite{Haider2006}, for individuals with different income growth there will nevertheless exist an age $t^*$ around midlife at which the expected difference between individuals' log annual incomes equals the expected difference between their lifetime incomes.}  The key implication is that researchers can reduce lifecycle bias by measuring income at mid-age.

As shown in Table \ref{tab:geiv}, this generalization of the classical error-in-variables model captures the relation between annual and lifetime incomes remarkably well. The insight that $\lambda_{st}$ increases over age and approximates one around mid-age holds in simulated income data calibrated to the US labor market (based on \citealt{Guvenen2009}, details available upon request), as well as in actual income series from Sweden and the US. However, the approach is subject to some limitations. First, lifecycle bias may not be fully eliminated at the age at which $\lambda_{st}=1$ because the assumption $Cov(y^*_{fi}, u_{sit})=0$ tends to be violated if income growth varies with parental background even conditional on a child's own lifetime income (as indicated by Figure \ref{fig:fig1} and shown formally in \citealt{Nybom2016}). 

Second, the optimal age $t^*$  at which $\lambda_{st}=1$ is rarely known, as its estimation requires data on lifetime incomes. In practice, applications follow instead a simple rule-of-thumb to measure income at \textit{some} point in midlife. Yet \cite{Haider2006} warn that $t^*$ is likely to vary across countries, and Table \ref{tab:geiv} shows that even slight deviations from this optimal age yield substantially different estimates. The rule-of-thumb estimates prevalent in the literature may therefore contain large biases, in particular given the extent to which the age at measurement varies across studies (see Table \ref{tab:literature}). 

\begin{table}[h!]
\centering
\caption{Lifecycle Bias and the Generalized-Errors-in-Variables Model} 
    \begin{tabular}{cccccc}
    \toprule
    \toprule
    \multicolumn{3}{c}{Swedish Register Data} & \multicolumn{3}{c}{US Simulated Data} \\
    \midrule
    Son's Age & $\lambda_{st}$ & $\beta_{t}$ & Son's Age & $\lambda_{st}$ & $\beta_{t}$\\
    \midrule
    33    & 0.858 & 0.221 & 41    & 0.896 & 0.461 \\
    34    & 0.913 & 0.237 & 42    & 0.958 & 0.470 \\
    35    & 0.969 & 0.253 & 43    & 0.997 & 0.506 \\
    36    & 1.024 & 0.270 & 44    & 1.036 & 0.518 \\
    37    & 1.080 & 0.285 & 45    & 1.047 & 0.525 \\
    True  &       & 0.253 &       &       & 0.497 \\
    \bottomrule
    \bottomrule
    \end{tabular}%
 \label{tab:geiv}
 \smallskip
\fnote{\textbf{Notes:} Estimates of $\lambda_{st}$ are based on equation \eqref{eq:eq1}. Estimates of $\beta_{t}$ are based on a regression of parental lifetime income on offspring annual income at age \textit{t}. Source: Swedish register data and simulated income data for the US, based on \cite{Guvenen2009}.}
\end{table}%

A third problem is that income around midlife is often not observed in the sample of interest. By definition, it will not be available if our interest centers on recent cohorts, who are still in their 20s or early 30s. \cite{Lee2009} provide an extension of the GEiV model for the estimation of mobility trends, which allows for the inclusion of observations at younger age by accounting for the age difference to a reference age. Lifecycle bias would not affect the estimated mobility trend if that bias remained sufficiently stable (i.e., if $\lambda_{st}$ and $Cov(y^*_{fi}, u_{sit})$ remain constant) over cohorts.  However, the structure of income profiles does change over time (e.g., \citealt{Guvenen2009}), and the age profile of $\lambda$ varies over the cohorts in our benchmark sample: at age 35, estimates of $\lambda$ vary between 1 and 1.2 between cohorts born in the early vs. late 1950s, scaling estimates of the IGE accordingly.\footnote{This observation may reflect that income distributions, and therefore the value of $\lambda$, can change substantially with macroeconomic conditions – such as the recession that Sweden experienced in the early 1990s.} As a result, the IGE appears to increase twice as much when using incomes at age 35 rather than lifetime incomes (see Figure \ref{fig:HertzTrend}). These observations suggest that estimates based on a fixed age window or fixed reference age, while useful for identifying sudden or large shifts in mobility, might not provide a good approximation for more gradual mobility trends over time.

\setcounter{table}{0}
\renewcommand{\thetable}{C.\arabic{table}}
\setcounter{figure}{0}
\renewcommand{\thefigure}{C.\arabic{figure}}

\section{Modelling the Income Process: Fixed Effects}\label{app:hertz}

\begin{figure}[h!]
 \centering
\caption{\label{fig:hertz_vogel}{Illustration of Potential Problems with Fixed Effect Estimators}}
   \includegraphics[width=1\textwidth]{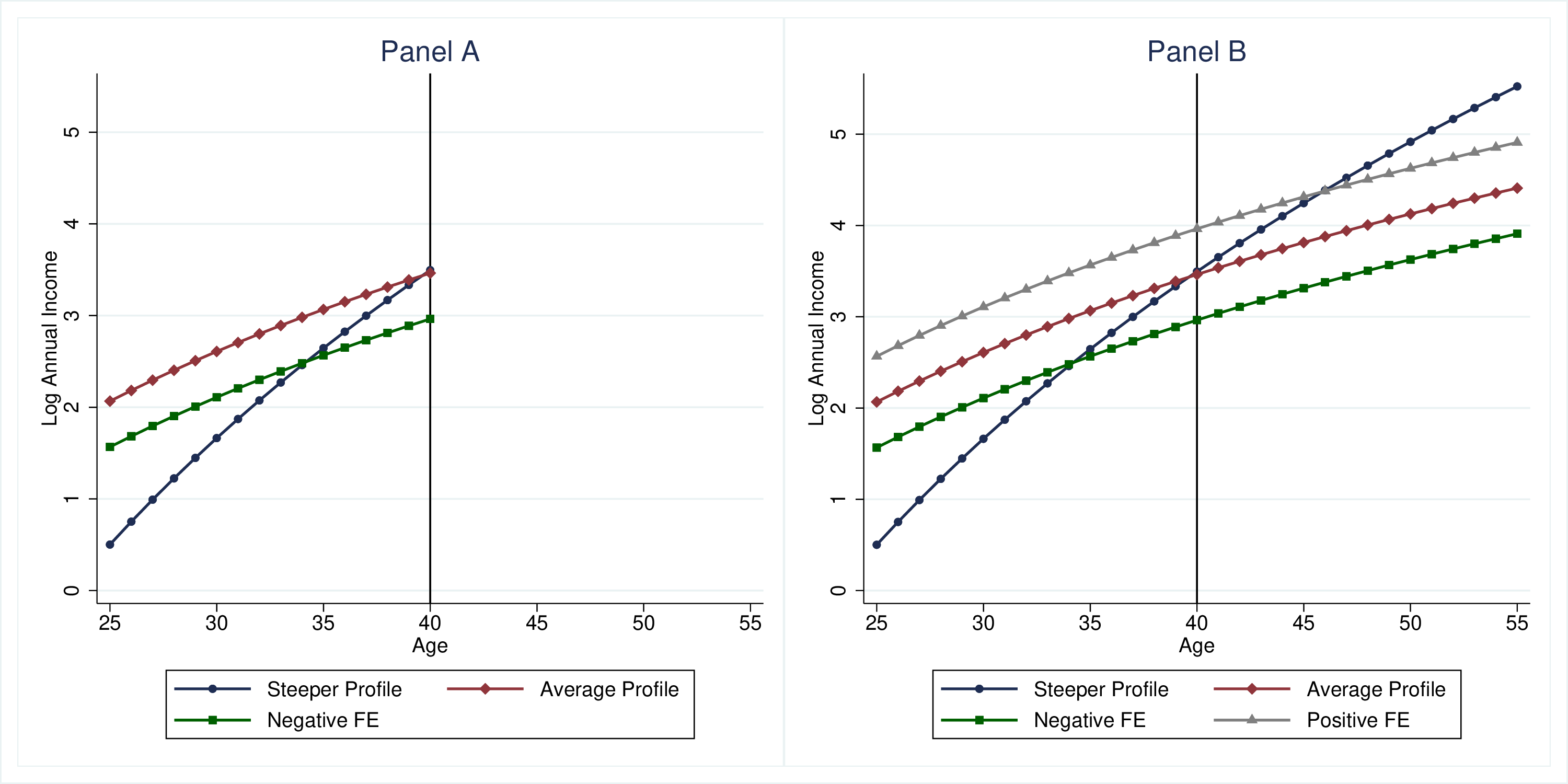}
   \fnote{\textbf{Notes:} In the Figure, the red line represents the average income profile in the population while the blue line represents the income trajectory of an individual with steeper than average profile. The gray and green lines correspond to the red line shifted by a positive or negative fixed effect, respectively.}
\end{figure}

Figure \ref{fig:hertz_vogel} provides intuition for why mobility estimates based on equation (\ref{eqvogelhertz}) remain sensitive to the age at which incomes are measured. Suppose the blue line (round dots) is the true income trajectory of individual \textit{i} with a steeper than average profile, while the red line (diamonds) is the average income profile in the population. Now, suppose we only observe incomes at age 25-40, as in Panel A of the figure. In this case, the predicted income profile for individual \textit{i} is given by the green profile (squares), corresponding to the red line plus a negative individual fixed effect. We would therefore understate the lifetime income of those with steeper profiles. Because income growth increases systematically with parental income even after conditioning on own education or occupation (see Section \ref{sec:incprocess}), the intergenerational elasticity is understated as well. The shorter and earlier the age range, the more we are understating the elasticity, as illustrated in Table \ref{tab:hertzvogeltab}.

The problem will be compounded when using equation (\ref{eqvogelhertz}) to predict lifetime incomes for both the child and the parent generation. Panel B of Figure  \ref{fig:hertz_vogel} illustrates why the the approach understates the income of children (observed early in life) with steeper than average profiles, and overstate the lifetime income of parents with steeper than average profiles (observed late in life). Suppose that the income profile of both parent and child is given by the red line, but that we observe the child earlier in life (e.g., ages 25-40) and the parent later in life (ages 40+). As individual heterogeneity can only be captured by the fixed effects, the father will have a positive fixed effect and the child a negative fixed effect. As a consequence, we would be understating the lifetime income of sons who have steeper than average profiles (green line), overstating the lifetime income of their fathers (gray line), and therefore, understating the intergenerational elasticity. 

\begin{figure}[h!]
 \centering
  \caption{\label{fig:HertzTrend}{Estimation of Trends in the IGE}}
   \includegraphics[width=1\textwidth]{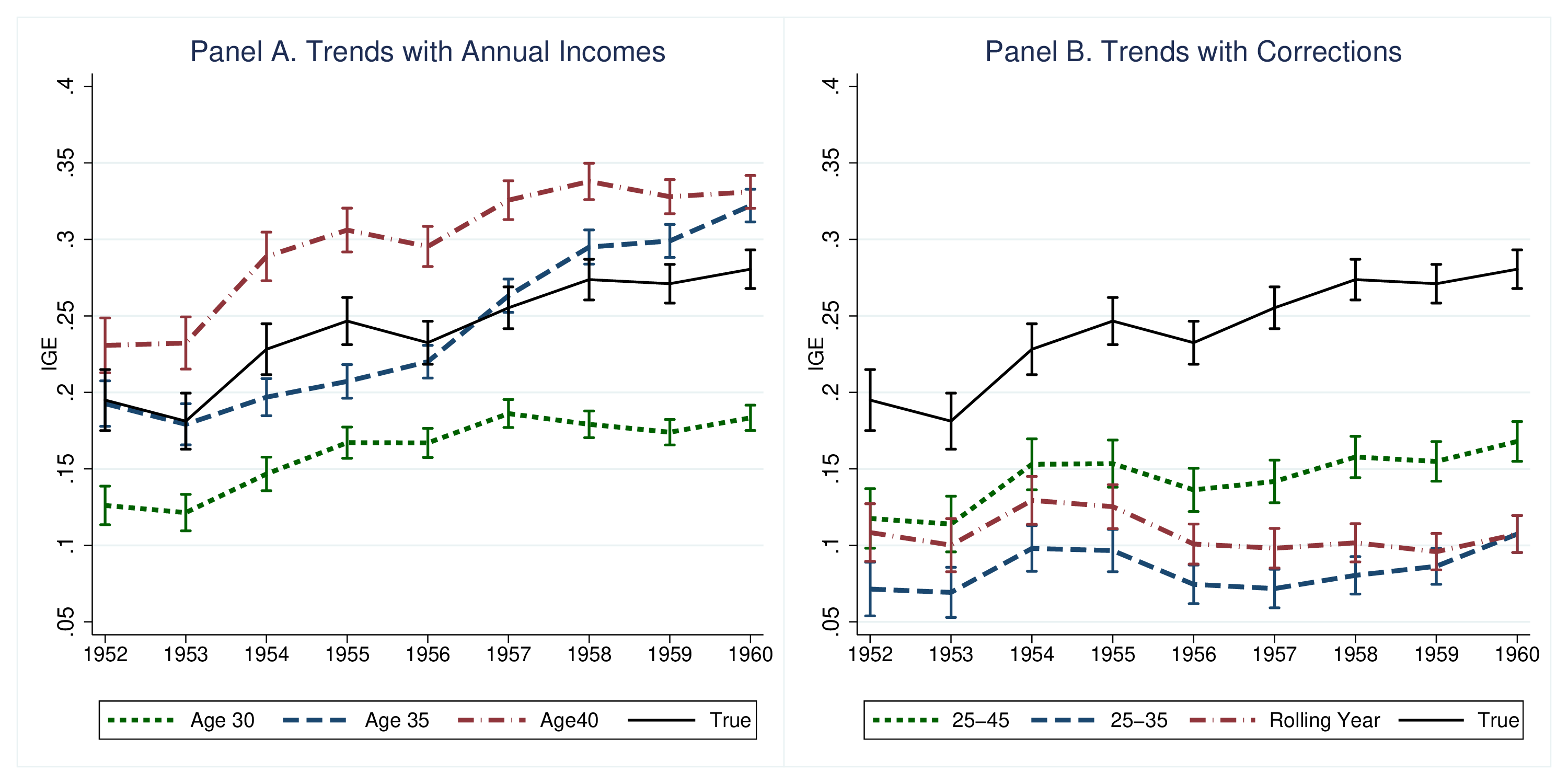}
   \fnote{\textbf{Notes:} In this Figure, we plot trends in the IGE using the Swedish Register data. In Panel A, we plot true trends using son's lifetime income (in black) and trends using annual incomes at ages 30, 35 or 40. In Panel B, we plot estimates of the IGE based on the first-step equation \eqref{eqvogelhertz} using either ages 25-45 or 25-35, to then predict income at age 25 for estimation of the IGE. For the line red in Panel B (``Rolling Year'') we instead use a rolling age window that reduces as the cohorts become younger (similar to \citealt{Hertz2007}). For the 1952 cohort the age range for estimation is 25-43, for the 1953 cohort the range is 25-42, and so on.}
\end{figure}

The same problem also affects the estimation of mobility trends, as is illustrated in Figure \ref{fig:HertzTrend} based on our Swedish benchmark sample. The ``true'' cohort trend (based on lifetime incomes, black line) increases for cohorts born in the 1950s. Panel A compares this benchmark to estimates based on income at a fixed age. While these estimators agree on the direction of the trend, the magnitudes differ. Panel B compares the benchmark to the two-step estimator as described in Section \ref{subsec:modellingprocess}. The trend is relatively well captured if the estimator is based on fixed age windows (blue and green lines). However, using a rolling age window – considering age 25-43 for the 1952 cohort but reducing the age range for more recent cohorts –  we fail to capture the increase in the IGE (red line). Our findings therefore suggest that trend estimates based on rolling age windows are susceptible to lifecycle effects.

\setcounter{table}{0}
\renewcommand{\thetable}{D.\arabic{table}}
\setcounter{figure}{0}
\renewcommand{\thefigure}{D.\arabic{figure}}

\section{Modelling the Income Process: Relating Growth and Levels}\label{app:creedy}

\cite{Creedy1988} proposes a correction method based on the insight that the dispersion of earnings tends to increase over age, even conditional on education or occupation. To account for this pattern, he assumes that income growth varies with the income \textit{rank} of the individual in the income distribution. An important advantage of this method is that it can be implemented in cross-sectional data sources. In a first step, we estimate how the mean and the variance of log income vary over age within each occupational or education group. Following \cite{Creedy1988}, we estimate
\begin{equation} \label{eq:creedy1}
y_{ij} = \beta_0 + \beta_1age_{ij}+ \beta_2age_{ij}^2+u_{ij},  
\end{equation}
separately by each occupational or education group \textit{j}, where $y_{ij}$ is the log income of individual \textit{i} and group \textit{j}. Then, we predict $\hat{\mu}_{tj}$, which is the average income by each occupational group \textit{j} and age group \textit{t}. The variance of log income $\sigma^{2}_{tj}$ is also computed within each group. Next, we estimate: 
\begin{equation} \label{eq:creedy2}
\sigma^{2}_{tj} = \beta_0 + \beta_1age_{tj} +\epsilon_{tj}, 
\end{equation}
and obtain predicted values for $\hat{\sigma}^{2}_{tj}$. Alternatively, one can obtain these measures from external sources. 

In a second step, these predicted values are used to rescale individual incomes to a common base year. First, compute the \textit{standardized} value of an individual's log-earnings,
\begin{equation}
\label{eq:CreedyStandardization}
z_t = y_t - \hat{\mu}_{tj}/\hat{\sigma}_{tj}.  
\end{equation}
Then, rescale these standardized incomes according to the occupation or education-specific age-earning profile to compute adjusted log earnings at a common age \textit{t*}: 
\begin{equation}
\label{eq:CreedyRescaling}
y_{t^{*}} = \hat{\mu}_{t^{*}j}+z\hat{\sigma}_{t^{*}j}.  
\end{equation}
Those adjusted earnings depend on a single observable income at age $t$ and on the values of $\hat{\mu}_t$ and $\hat{\sigma}_t$ that were predicted within the education and/or occupational group. Finally, we have adjusted income observations for different ages, computed based on a single cross-section observation and scaling factors. \cite{Creedy1988} proposes to either use adjusted earnings directly or to compute an aggregated discounted lifetime earnings measure for the estimation of the IGE. 

\begin{figure}[h!]
 \centering
 \caption{\label{fig:CreedyPlot}{Extrapolating from Observable Profiles}}
   \includegraphics[width=0.6\textwidth]{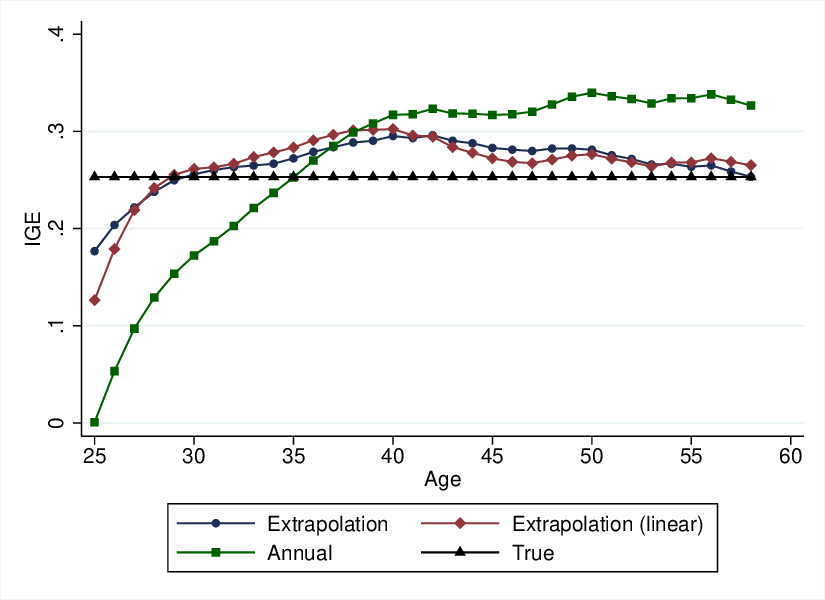}
    \fnote{\textbf{Notes:} 
    We compare IGE estimates based on the ``true'' lifetime income of sons (black line), estimates based on annual incomes (green line), and two versions of Creedy's proposed estimator. In the first, we approximate the profiles of $\hat{\mu}_{tj}$ and $\hat{\sigma}_{tj}^2$  with a linear function in age (red line). In the second, we use their non-parametric age profile as observed in the sample (blue line).}
\end{figure}

We implement this method in the Swedish data. We combine the first-step estimates of $\hat{\mu}_{tj}$ and $\hat{\sigma}_{tj}^2$  with an individual's earning at a certain age, to obtain his predicted income from ages 25-53 (which are then used to construct lifetime incomes). We therefore obtain a different measure of lifetime income, and a different estimate for the IGE, depending on the age at which we measure sons' income. We plot the resulting estimates of the IGE in Figure \ref{fig:CreedyPlot}. We compare estimates based on the ``true'' lifetime income of sons (black line), estimates based on annual incomes (green line), and two versions of Creedy's proposed estimator. In the first, we approximate the profiles of $\hat{\mu}_{tj}$ and $\hat{\sigma}_{tj}^2$  with a linear function in age (red line). In the second, we use their non-parametric age profile as observed in the sample (blue line).

The comparison demonstrates that estimates of the IGE can be significantly improved by taking the dispersion of income growth over age into account. The corrected estimates are within 20 percent of the benchmark over the age range 30 to 50, even if the age profiles of  $\hat{\mu}_{tj}$ and $\hat{\sigma}_{tj}^2$  are approximated linearly. The correction works less well than a correction based on the generalized errors-in-variables model proposed by \cite{Haider2006}, but it is also based on less stringent requirements – only the age pattern of the variances and means is required. As \cite{Creedy1988} discusses, the statistics that are necessary for the correction can potentially be estimated from a single cross-section. However, Figure \ref{fig:CreedyPlot} also shows that the correction method works only imperfectly, and tends to overstate the IGE over most of the age range. 

A key limitation is that equation (\ref{eq:CreedyRescaling}) rescales incomes based on the assumption that individual's rank in the widening income distribution remains stable over age: that is, individuals with high annual rank are assumed to have higher income growth in the future. This is not the case in practice, as is illustrated in Panel B of Figure \ref{fig:fig1}. Because of short-term variability, annual incomes are instead mean-reverting – individuals with high income rank at age $t$ tend to have lower income growth in the next few years. By not accounting for this mean-reverting influence, the imputation in equation (\ref{eq:CreedyRescaling}) tends to overstate the variance of lifetime incomes and therefore the IGE. The method performs better the more important the heterogeneous growth rates are compared to the transitory shock component. For example, in the HIP process proposed by \cite{Guvenen2009}, incomes at early ages are dominated by transitory shocks from an AR(1) process (and intercepts), while incomes at later ages are dominated by idiosyncratic growth rates, and the extrapolation from observed ranks would work better at later ages.

\setcounter{table}{0}
\renewcommand{\thetable}{E.\arabic{table}}
\setcounter{figure}{0}
\renewcommand{\thefigure}{E.\arabic{figure}}

\section{Levels vs Logs in the First Step of the Estimator}\label{app:levelsappendix}

The first step of our proposed lifecycle estimator consists of the estimation of variants of the following equation, as explained in Section \ref{sec:lifecyclestimator}:

\begin{equation} \label{eqtwostepapp}
y_{ict} = \alpha_i + g(A_{ict},Z_{ic}) + f(A_{ict},Z_{ic},P_{ic}) + \varepsilon_{ict}.
\end{equation}

In this section, we set $y_{ict}$ to be the absolute income of individual $i$ from cohort $c$ in period $t$ instead of log income, our preferred specification. There are advantages and disadvantages of estimating equation (\ref{eqtwostepapp}) with absolute instead of log incomes. The main advantage is the avoidance of the re-transformation problem described in Section \ref{sec:lifecyclestimator}. 
While the fitted values from the estimation of equation (\ref{eqtwostepapp}) with logs have mean zero by construction ($E[\hat{\varepsilon}_{ict}]=0$), their mean will be positive after transformation to absolute income ($E[exp(\hat{\varepsilon}_{ict})]>0$) before aggregation of lifetime incomes. In practice, this implies that we need to add an additional step in our estimation. If incomes are estimated and predicted in levels, we are not subject to the re-transformation problem. However, other challenges become evident, as discussed below. 

Table \ref{tab:lifecycleestimatorlevels} shows the performance of the lifecycle estimators with absolute incomes in the first step instead of log incomes for the Swedish Registers. The first noteworthy difference between Table \ref{tab:lifecycleestimatorlevels} and the benchmark Table \ref{tab:lifecycleestimator} (with log incomes) is the lower number of observations for each age range, especially in column (7). This happens because some annual predictions with absolute incomes happen to be negative. Thus, when we take logs for the computation of lifetime incomes, these observations are assigned a missing value. This is particularly problematic for the ``slope-level'' estimator of column (7). 

To overcome this issue, we bottom code predictions of annual incomes to 10,000 SEK (around 912 USD). This is the same bottom coding done to the actual annual incomes.  Table \ref{tab:lifecycleestimatorlevelstrim} shows the performance of the lifecycle estimators with absolute incomes in the first step and with predicted annual incomes bottom-coded to 10,000 SEK. After the bottom coding, we obtain the exact number of observations as in the benchmark Table \ref{tab:lifecycleestimator}. The baseline lifecycle estimator of Table \ref{tab:lifecycleestimatorlevelstrim} column (3) shows a clear lifecycle pattern, but lower levels for most age ranges (except the first) when compared to Table \ref{tab:lifecycleestimator}. The parental lifecycle estimators of columns (4) and (5) and the slope-level estimator of column (7) are very similar to the ones in Table \ref{tab:lifecycleestimator}, with the exception of the lowest age range, where the estimator of Table \ref{tab:lifecycleestimatorlevelstrim} performs better. Finally, the parental quadratic estimator without fixed effects (column 6) is slightly lower than the one in Table \ref{tab:lifecycleestimator} and the benchmark of column (1), although still performing well.

Table \ref{tab:lifecycleestimatorlevelspsid} shows the comparison between estimates in logs and levels for the PSID. The performance of the parental lifecycle estimators in levels (Table \ref{tab:lifecycleestimatorlevelspsid}, columns 4 and 5) is similar to the one of the estimators in logs (Table \ref{tab:lifecyclepsid}). The exception is the 22-27 age range, where the estimator in levels is closer to the benchmark. In turn, the slope-level estimator in levels (column 7) of age ranges 22-27 and 22-30 performs worse than the one in logs, overstating the IGE compared to the benchmark. 

In sum, our lifecycle estimators show good performance regardless of whether we estimate the first step with log or absolute annual incomes. With log annual incomes, we need to correct for the re-transformation problem. With absolute annual incomes, we need to take into account negative predictions, which might exclude some observations entirely. We suggest that practitioners experiment with both approaches and choose the one that best fits their data and application. 

\begin{table}[h!]
\centering
       \caption{The Lifecycle Estimator - First Step in Levels (Swedish Registers)}
     \begin{adjustbox}{width=\textwidth}
         \begin{tabular}{lcccccccc}
    \toprule
    \toprule
          & \multicolumn{2}{c}{Direct estimator} &       & \multicolumn{5}{c}{Lifecycle estimator} \\
\cmidrule{2-3}\cmidrule{5-9}          & Lifetime & Annual &       & Baseline  & Parental & Parental & Parental & Slope-level \\
          &       &       &       &       & Linear & Quadratic & Quadratic & Quadratic \\
              &       &       &       &   FE  & FE & FE & no FE & FE \\
    Son's Age & (1)   & (2)   &       & (3)   & (4)   & (5)   & (6)   & (7)\\
    \midrule
    Age 25-27 & 0.252 & 0.045 &       & 0.151 & 0.240 & 0.280 & 0.233 & 0.158 \\
          & (0.004) & (0.002) &       & (0.002) & (0.002) & (0.002) & (0.001) & (0.004) \\
    N=    & 94,098 & 281,504 &       & 94,098 & 94,098 & 94,098 & 94,098 & 90,073 \\
    $R^2$ & 0.049 & 0.001 &       & 0.053 & 0.124 & 0.162 & 0.385 & 0.020 \\
          &       &       &       &       &       &       &       &  \\
    Age 25-30 & 0.253 & 0.100 &       & 0.162 & 0.233 & 0.272 & 0.232 & 0.199 \\
          & (0.004) & (0.002) &       & (0.002) & (0.002) & (0.002) & (0.001) & (0.003) \\
    N=    & 94,192 & 561,502 &       & 94,192 & 94,192 & 94,192 & 94,192 & 91,353 \\
    $R^2$ & 0.050 & 0.006 &       & 0.065 & 0.124 & 0.164 & 0.385 & 0.034 \\
          &       &       &       &       &       &       &       &  \\
    Age 25-35 & 0.254 & 0.158 &       & 0.181 & 0.234 & 0.255 & 0.232 & 0.221 \\
          & (0.004) & (0.001) &       & (0.002) & (0.002) & (0.002) & (0.001) & (0.004) \\
    N=    & 94,262 & 1,024,560 &       & 94,262 & 94,262 & 94,262 & 94,262 & 91,176 \\
   $R^2$ & 0.050 & 0.013 &       & 0.071 & 0.113 & 0.132 & 0.384 & 0.041 \\
          &       &       &       &       &       &       &       &  \\
    Age 25-40 & 0.254 & 0.204 &       & 0.198 & 0.250 & 0.254 & 0.231 & 0.239 \\
          & (0.004) & (0.001) &       & (0.002) & (0.002) & (0.002) & (0.001) & (0.004) \\
    N=    & 94,310 & 1,482,460 &       & 94,310 & 94,310 & 94,310 & 94,310 & 91,790 \\
    $R^2$ & 0.050 & 0.018 &       & 0.071 & 0.108 & 0.111 & 0.383 & 0.047 \\
          &       &       &       &       &       &       &       &  \\
    Age 25-45 & 0.254 & 0.234 &       & 0.212 & 0.267 & 0.255 & 0.232 & 0.244 \\
          & (0.004) & (0.001) &       & (0.003) & (0.003) & (0.003) & (0.001) & (0.004) \\
    N=    & 94,323 & 1,935,127 &       & 94,323 & 94,323 & 94,323 & 94,323 & 92,386 \\
    $R^2$ & 0.050 & 0.021 &       & 0.069 & 0.105 & 0.097 & 0.385 & 0.049 \\
   \bottomrule
    \bottomrule
    \end{tabular}%
   \end{adjustbox}
   \smallskip
  \fnote{\textbf{Notes:} The table reports the slope coefficient from a regression of son's income on father's lifetime income. The measure for son's income is log lifetime income in column (1), the pooled log annual incomes from age 25 to the indicated upper age bound in column (2), or the predicted lifetime income from a first-step estimation in the indicated age range of equation \eqref{eqtwostep} in columns (3)-(6) or equation \eqref{eqtwostepFE} in column (7). See text for detailed definitions of each estimator. Robust standard errors in parentheses.}
  \label{tab:lifecycleestimatorlevels}%
\end{table}%

\begin{table}[h!]
\centering
      \caption{The Lifecycle Estimator - First Step in Levels with Bottom Coding (Swedish Registers)}
     \begin{adjustbox}{width=\textwidth}
         \begin{tabular}{lcccccccc}
    \toprule
    \toprule
          & \multicolumn{2}{c}{Direct estimator} &       & \multicolumn{5}{c}{Lifecycle estimator} \\
\cmidrule{2-3}\cmidrule{5-9}          & Lifetime & Annual &       & Baseline  & Parental & Parental & Parental & Slope-level \\
          &       &       &       &       & Linear & Quadratic & Quadratic & Quadratic \\
              &       &       &       &   FE  & FE & FE & no FE & FE \\
    Son's Age & (1)   & (2)   &       & (3)   & (4)   & (5)   & (6)   & (7)\\ 
    \midrule
    Age 25-27 & 0.253 & 0.046 &       & 0.151 & 0.239 & 0.277 & 0.233 & 0.227 \\
          & (0.004) & (0.002) &       & (0.002) & (0.002) & (0.002) & (0.001) & (0.005) \\
    N=    & 94,100 & 281,510 &       & 94,100 & 94,100 & 94,100 & 94,100 & 94,100 \\
    $R^2$ & 0.050 & 0.001 &       & 0.054 & 0.124 & 0.161 & 0.385 & 0.026 \\
          &       &       &       &       &       &       &       &  \\
    Age 25-30 & 0.253 & 0.101 &       & 0.162 & 0.232 & 0.271 & 0.232 & 0.251 \\
          & (0.004) & (0.002) &       & (0.002) & (0.002) & (0.002) & (0.001) & (0.004) \\
    N=    & 94,194 & 561,514 &       & 94,194 & 94,194 & 94,194 & 94,194 & 94,194 \\
    $R^2$ & 0.050 & 0.006 &       & 0.065 & 0.125 & 0.164 & 0.385 & 0.039 \\
          &       &       &       &       &       &       &       &  \\
    Age 25-35 & 0.254 & 0.159 &       & 0.180 & 0.234 & 0.255 & 0.232 & 0.267 \\
          & (0.004) & (0.001) &       & (0.002) & (0.002) & (0.002) & (0.001) & (0.004) \\
    N=    & 94,264 & 1,024,582 &       & 94,264 & 94,264 & 94,264 & 94,264 & 94,264 \\
    $R^2$ & 0.050 & 0.013 &       & 0.071 & 0.113 & 0.133 & 0.384 & 0.047 \\
          &       &       &       &       &       &       &       &  \\
    Age 25-40 & 0.254 & 0.204 &       & 0.197 & 0.249 & 0.253 & 0.231 & 0.263 \\
          & (0.004) & (0.001) &       & (0.002) & (0.002) & (0.002) & (0.001) & (0.004) \\
    N=    & 94,311 & 1,482,461 &       & 94,311 & 94,311 & 94,311 & 94,311 & 94,311 \\
    $R^2$ & 0.050 & 0.018 &       & 0.071 & 0.109 & 0.112 & 0.383 & 0.052 \\
          &       &       &       &       &       &       &       &  \\
    Age 25-45 & 0.254 & 0.234 &       & 0.210 & 0.266 & 0.255 & 0.232 & 0.257 \\
          & (0.004) & (0.001) &       & (0.002) & (0.002) & (0.002) & (0.001) & (0.004) \\
    N=    & 94,339 & 1,935,194 &       & 94,339 & 94,339 & 94,339 & 94,339 & 94,339 \\
    $R^2$ & 0.050 & 0.021 &       & 0.070 & 0.108 & 0.100 & 0.385 & 0.054 \\
    \bottomrule
    \bottomrule
    \end{tabular}%
  \end{adjustbox}
  \smallskip
  \fnote{\textbf{Notes:} The table reports the slope coefficient from a regression of son's income on father's lifetime income. The measure for son's income is log lifetime income in column (1), the pooled log annual incomes from age 25 to the indicated upper age bound in column (2), or the predicted lifetime income from a first-step estimation in the indicated age range of equation \eqref{eqtwostep} in columns (3)-(6) or equation \eqref{eqtwostepFE} in column (7). See text for detailed definitions of each estimator. Robust standard errors in parentheses.}
  \label{tab:lifecycleestimatorlevelstrim}%
\end{table}%

\begin{table}[h!]
\centering
       \caption{The Lifecycle Estimator - First Step in Levels (PSID)}
     \begin{adjustbox}{width=\textwidth}
         \begin{tabular}{lcccccccc}
    \toprule
    \toprule
          & \multicolumn{2}{c}{Direct estimator} &       & \multicolumn{5}{c}{Lifecycle estimator} \\
\cmidrule{2-3}\cmidrule{5-9}          & Lifetime & Annual &       & Baseline  & Parental & Parental & Parental & Slope-level \\
          &       &       &       &       & Linear & Quadratic & Quadratic & Quadratic \\
              &       &       &       &   FE  & FE & FE & no FE & FE \\
    Son's Age & (1)   & (2)   &       & (3)   & (4)   & (5)   & (6)   & (7)\\
\cmidrule{1-3}\cmidrule{5-9}    Age 22-27 & 0.426 & 0.242 &       & 0.303 & 0.399 & 0.451 & 0.432 & 0.506 \\
    N=892 & (0.035) & (0.016) &       & (0.024) & (0.024) & (0.024) & (0.014) & (0.045) \\
    $R^2$ & 0.142 & 0.040 &       & 0.152 & 0.238 & 0.288 & 0.513 & 0.123 \\
          &       &       &       &       &       &       &       &  \\
    Age 22-30 & 0.426 & 0.280 &       & 0.311 & 0.394 & 0.426 & 0.432 & 0.514 \\
    N=892 & (0.035) & (0.013) &       & (0.024) & (0.024) & (0.024) & (0.014) & (0.045) \\
    $R^2$ & 0.142 & 0.053 &       & 0.163 & 0.237 & 0.268 & 0.513 & 0.129 \\
          &       &       &       &       &       &       &       &  \\
    Age 22-35 & 0.426 & 0.321 &       & 0.339 & 0.402 & 0.417 & 0.432 & 0.479 \\
    N=892 & (0.035) & (0.011) &       & (0.025) & (0.025) & (0.025) & (0.014) & (0.041) \\
    $R^2$ & 0.142 & 0.064 &       & 0.166 & 0.220 & 0.235 & 0.513 & 0.135 \\
          &       &       &       &       &       &       &       &  \\
    Age 22-40 & 0.426 & 0.354 &       & 0.372 & 0.443 & 0.442 & 0.432 & 0.454 \\
    N=892 & (0.035) & (0.010) &       & (0.028) & (0.028) & (0.028) & (0.014) & (0.038) \\
    $R^2$ & 0.142 & 0.070 &       & 0.166 & 0.222 & 0.220 & 0.513 & 0.137 \\
          &       &       &       &       &       &       &       &  \\
    Age 22-45 & 0.426 & 0.373 &       & 0.384 & 0.440 & 0.423 & 0.432 & 0.427 \\
    N=892 & (0.035) & (0.009) &       & (0.030) & (0.030) & (0.030) & (0.014) & (0.036) \\
    $R^2$ & 0.142 & 0.073 &       & 0.156 & 0.196 & 0.183 & 0.513 & 0.136 \\
    \bottomrule
    \bottomrule
    \end{tabular}%
  \end{adjustbox}
  \smallskip
  \fnote{\textbf{Notes:} The table reports the slope coefficient from a regression of son's income on father's lifetime income. The measure for son's income is log lifetime income in column (1), the pooled log annual incomes from age 25 to the indicated upper age bound in column (2), or the predicted lifetime income from a first-step estimation in the indicated age range of equation \eqref{eqtwostep} in columns (3)-(6) or equation \eqref{eqtwostepFE} in column (7). See text for detailed definitions of each estimator. Robust standard errors in parentheses.}
  \label{tab:lifecycleestimatorlevelspsid}%
\end{table}%

\clearpage
\pagebreak
\newpage
\setcounter{table}{0}
\renewcommand{\thetable}{F.\arabic{table}}
\setcounter{figure}{0}
\renewcommand{\thefigure}{F.\arabic{figure}}

\section{Supplementary Evidence on Cohort Trends}\label{app:supp_trends}

\begin{table}[h!]
\caption{Tests of trends estimates (Swedish Data)}
   \begin{adjustbox}{width=\textwidth}
\begin{tabular}{lcccccc}
\toprule
\toprule
 & \multicolumn{6}{c}{Lifecycle estimator, different specifications}\tabularnewline
 & (1) & (2) & (3) & (4) & (5) & (6)\tabularnewline
\cmidrule{2-7}
Cohorts & 0.196 & 0.197 & 0.195 & 0.197 & 0.197 & 0.196\tabularnewline
1950-59 & (0.002) & (0.002) & (0.002) & (0.002) & (0.002) & (0.002)\tabularnewline
 &  &   &   &   &   &  \tabularnewline
Cohorts & 0.190 & 0.206 & 0.209 & 0.208 & 0.208 & 0.213\tabularnewline
1960-69 & (0.002) & (0.002) & (0.002) & (0.002) & (0.002) & (0.002)\tabularnewline
 &   &   &   &   &   &  \tabularnewline
Cohorts & 0.168 & 0.200 & 0.212 & 0.201 & 0.200 & 0.200\tabularnewline
1970-79 & (0.002) & (0.002) & (0.002) & (0.002) & (0.002) & (0.002)\tabularnewline
 &   &   &   &   &   &  \tabularnewline
Cohorts & 0.134 & 0.186 & 0.215 & 0.181 & 0.183 & 0.164\tabularnewline
1980-89 & (0.002) & (0.002) & (0.002) & (0.002) & (0.002) & (0.002)\tabularnewline
\hline
Education X age &  & X &  & X & X & X\tabularnewline
Parent educ. X age &  & X &  & X & X & X\tabularnewline
Parent income X age &  &  & X &  &  & X\tabularnewline
Educ.X age X cohort &  &   &  & X &  & \tabularnewline
Parent educ.X age X cohort &  &  &  &  & X & \tabularnewline
Parent inc.X age X cohort &  &  &  &  &  & X\tabularnewline
\midrule
$R^{2}$ & 0.050 & 0.061 & 0.065 & 0.060 & 0.060 & 0.059\tabularnewline
Individuals & 1,844,829 & 1,844,829 & 1,844,829 & 1,844,829 & 1,844,829 & 1,844,829\tabularnewline
\bottomrule
\bottomrule
\end{tabular}
  \end{adjustbox}
  \label{tab:trends_predictors_Swe}%
  \smallskip
 \fnote{\textbf{Notes:} The table reports different variants of the lifecycle estimator for cohort trends in Sweden, in which the set of first-step regressors varies across columns. The specifications in all columns include individual fixed effects, year fixed effects, and a standardized age profile interacted with cohort dummies. Column (2) adds quadratic age profiles that are allowed to vary by own and parental education. Column (3) instead includes an interaction between quadratic age profiles and fathers' income. We next add interactions between cohort dummies, a standardized profile, and either own education (column 4), parental education (column 5), or parental income (column 6).} 
\end{table}

\begin{table}[h!]
\caption{Tests of trends estimates (PSID)}
 \begin{adjustbox}{width=\textwidth}
\begin{tabular}{lcccccc}
\toprule
\toprule

 & \multicolumn{6}{c}{Lifecycle estimator, different specifications}\tabularnewline
 & (1) & (2) & (3) & (4) & (5) & (6)\tabularnewline
\cmidrule{2-7}
Cohorts & 0.385 & 0.422 & 0.434 & 0.423 & 0.425 & 0.437\tabularnewline
1950-59 & (0.038) & (0.040) & (0.037) & (0.040) & (0.040) & (0.040)\tabularnewline
 &   &   &   &   &   &  \tabularnewline
Cohorts & 0.391 & 0.437 & 0.453 & 0.435 & 0.432 & 0.438\tabularnewline
1960-69 & (0.035) & (0.036) & (0.034) & (0.036) & (0.036) & (0.036)\tabularnewline
 &   &   &   &   &   &  \tabularnewline
Cohorts & 0.396 & 0.457 & 0.468 & 0.459 & 0.457 & 0.470\tabularnewline
1970-79 & (0.031) & (0.033) & (0.031) & (0.033) & (0.033) & (0.033)\tabularnewline
 &   &   &   &   &   &  \tabularnewline
Cohorts & 0.290 & 0.376 & 0.405 & 0.391 & 0.384 & 0.432\tabularnewline
1980-89 & (0.024) & (0.025) & (0.024) & (0.026) & (0.026) & (0.026)\tabularnewline
\hline 
Education X age &  & X &  & X & X & X\tabularnewline
Parent educ. X age &  & X &  & X & X & X\tabularnewline
Parent income X age &  &  & X &  &  & X\tabularnewline
Educ. X age X cohort &  &   &  & X &  & \tabularnewline
Parent educ. X age X cohort &  &  &  &  & X & \tabularnewline
Parent inc.X age X cohort &  &  &  &  &  & X\tabularnewline
\hline 
$R^{2}$ & 0.123 & 0.148 & 0.173 & 0.150 & 0.150 & 0.165\tabularnewline
Individuals & 4,939 & 4,937 & 4,939 & 4,937 & 4,937 & 4,937\tabularnewline
\bottomrule
\bottomrule
\end{tabular}
  \end{adjustbox}
  \label{tab:trends_predictors_PSID}
  \smallskip
 \fnote{\textbf{Notes:} The table reports different variants of the lifecycle estimator for cohort trends in the US, in which the set of first-step regressors varies across columns. The specifications in all columns include individual fixed effects, year fixed effects and a standardized age profile interacted with cohort dummies. Column (2) adds quadratic age profiles that are allowed to vary by own and parental education. Column (3) instead includes an interaction between quadratic age profiles and fathers' income. We next add interactions between cohort dummies, a standardized profile, and either own education (column 4), parental education (column 5), or parental income (column 6).} 
\end{table}

\begin{sidewaystable}[ph]
\centering 
 \caption{Parental income gradients in earnings growth over the lifecycle and
cohorts}
 \begin{adjustbox}{width=\textwidth}
\begin{tabular}{lcccccccccc}
\toprule
\toprule
          & \multicolumn{10}{c}{Earnings growth ages} \\
          & \multicolumn{2}{c}{Ages 25-30} & \multicolumn{2}{c}{Ages 30-35} & \multicolumn{2}{c}{Ages 35-40} & \multicolumn{2}{c}{Ages 40-45} & \multicolumn{2}{c}{Ages 45-50} \\
\cmidrule{2-11}    Cohort 1950-59 & \multicolumn{1}{l}{12.357***} & \multicolumn{1}{l}{4.139***} & \multicolumn{1}{l}{4.878***} & \multicolumn{1}{l}{1.767***} & \multicolumn{1}{l}{2.939***} & \multicolumn{1}{l}{0.558***} & \multicolumn{1}{l}{2.279***} & \multicolumn{1}{l}{1.080***} & \multicolumn{1}{l}{-0.156} & \multicolumn{1}{l}{-0.609***} \\
          & \multicolumn{1}{l}{(0.316)} & \multicolumn{1}{l}{(0.297)} & \multicolumn{1}{l}{(0.144)} & \multicolumn{1}{l}{(0.147)} & \multicolumn{1}{l}{(0.119)} & \multicolumn{1}{l}{(0.123)} & \multicolumn{1}{l}{(0.116)} & \multicolumn{1}{l}{(0.120)} & \multicolumn{1}{l}{(0.111)} & \multicolumn{1}{l}{(0.115)} \\
    Cohort 1960-69 & \multicolumn{1}{l}{10.718***} & \multicolumn{1}{l}{4.814***} & \multicolumn{1}{l}{4.326***} & \multicolumn{1}{l}{1.079***} & \multicolumn{1}{l}{1.228***} & \multicolumn{1}{l}{-0.108} & \multicolumn{1}{l}{0.147} & \multicolumn{1}{l}{-0.470***} & \multicolumn{1}{l}{-0.148} & \multicolumn{1}{l}{-0.441***} \\
          & \multicolumn{1}{l}{(0.159)} & \multicolumn{1}{l}{(0.151)} & \multicolumn{1}{l}{(0.127)} & \multicolumn{1}{l}{(0.130)} & \multicolumn{1}{l}{(0.108)} & \multicolumn{1}{l}{(0.111)} & \multicolumn{1}{l}{(0.100)} & \multicolumn{1}{l}{(0.102)} & \multicolumn{1}{l}{(0.102)} & \multicolumn{1}{l}{(0.105)} \\
    Cohort 1970-79 & \multicolumn{1}{l}{7.070***} & \multicolumn{1}{l}{2.405***} & \multicolumn{1}{l}{1.529***} & \multicolumn{1}{l}{-0.249*} & \multicolumn{1}{l}{0.566***} & \multicolumn{1}{l}{-0.280**} &       &       & \multicolumn{1}{l}{} & \multicolumn{1}{l}{} \\
          & \multicolumn{1}{l}{(0.149)} & \multicolumn{1}{l}{(0.143)} & \multicolumn{1}{l}{(0.103)} & \multicolumn{1}{l}{(0.105)} & \multicolumn{1}{l}{(0.095)} & \multicolumn{1}{l}{(0.097)} &       &       & \multicolumn{1}{l}{} & \multicolumn{1}{l}{} \\
    Cohort 1980-89 & \multicolumn{1}{l}{4.933***} & \multicolumn{1}{l}{1.242***} &       &       & \multicolumn{1}{l}{} & \multicolumn{1}{l}{} & \multicolumn{1}{l}{} & \multicolumn{1}{l}{} & \multicolumn{1}{l}{} & \multicolumn{1}{l}{} \\
          & \multicolumn{1}{l}{(0.134)} & \multicolumn{1}{l}{(0.127)} &       &       & \multicolumn{1}{l}{} & \multicolumn{1}{l}{} & \multicolumn{1}{l}{} & \multicolumn{1}{l}{} & \multicolumn{1}{l}{} & \multicolumn{1}{l}{} \\
    \midrule
    Education control &       & X     &       & X     &       & X     &       & X     &       & X \\
    N     & 1,108,129 & 1,108,129 & 1,198,547 & 1,198,546 & 1,143,144 & 1,143,144 & 936,700 & 936,700 & 712,634 & 712,634 \\
    $R^2$    & 0.027 & 0.130 & 0.016 & 0.041 & 0.010 & 0.020 & 0.001 & 0.005 & 0.002 & 0.003 \\
    \bottomrule 
\bottomrule
\end{tabular}
 \end{adjustbox}
  \label{tab:trends_growth_Swe}
 \fnote{\textbf{Notes:} The table reports coefficients from regressions of age-varying earnings growth on log father's income divided by 100, separately by cohort group. The dependent variable is the five-year change in log annual income (i.e., the difference between log income at age $t$ and log income at $t-5$), measured at $t=30, t=35, t=40, t=45, t=50$. This means that in the age range 25-30 the dependent variable would be the difference in log income between age 30 and age 25, for example. Every other column also controls for education level (distinguishing 15 levels). Empty cells by construction lack data for the indicated age range. Robust standard errors in parentheses; \sym{*} $p<0.05$, \sym{{*}{*}} $p<0.01$, \sym{{*}{*}{*}}
$p<0.001$}

\end{sidewaystable}
\clearpage
\pagebreak
\newpage

\setcounter{table}{0}
\renewcommand{\thetable}{G.\arabic{table}}
\setcounter{figure}{0}
\renewcommand{\thefigure}{G.\arabic{figure}}
\section{Machine Learning and Regularization} \label{app:machinelearning}

Ainteresting question is whether machine learning and regularization methods to select the first-step regressors ld improve the performance of the lifecycle estimator. To study this, we implemented this first step in the Swedish data using penalized regression methods, such as lasso or elastic nets. While we provide a short summary in the main text, we provide here a more extensive set of results.

\begin{table}[h!]
\centering
  \caption{ML Estimation of Lifecycle Profiles}
     \begin{adjustbox}{width=\textwidth}
         \begin{tabular}{lcccccccc}
    \toprule
    \toprule
          & \multicolumn{1}{c}{Direct} &       & \multicolumn{6}{c}{Lifecycle estimator} \\
\cmidrule{2-2}\cmidrule{4-9}          & Lifetime &       & Parental & Lasso & Lasso & Lasso & Lasso & Lasso \\
          &       &       &  Quadratic & $\lambda=0.01$ & $\lambda=0.001$ & $\lambda=0.0001$ & $\lambda=0.00001$ & $\lambda=0.01$ \\
          &       &       &    &  &  &  &  & not pen. \\        
    Son's Age & (1)  & & (2)  & (3)   & (4)   & (5)   & (6)   & (7) \\
    \midrule
Age 25-27 & 0.219 &  & 0.203 & 0.040 & 0.117 & 0.173 & 0.203 & 0.203 \\
N=71,794 & (0.003) &  & (0.004) & (0.004) & (0.004) & (0.004) & (0.004) & (0.004) \\
\# vars &  &  & 76 & 233 & 233 & 233 & 233 & 233 \\
\# vars selected &   &  & 76 & 16 & 48 & 105 & 174 & 42 \\[.1cm]
Age 25-30 & 0.219 &  & 0.241 & 0.099 & 0.191 & 0.239 & 0.254 & 0.268 \\
N=71,846 & (0.003) &  & (0.004) & (0.003) & (0.004) & (0.004) & (0.004) & (0.004) \\
\# vars &  &  & 76 & 234 & 234 & 234 & 234 & 234 \\
\# vars selected &  &  & 76 & 15 & 52 & 115 & 170 & 45 \\[.1cm]
Age 25-35 & 0.219 &  & 0.229 & 0.151 & 0.208 & 0.235 & 0.236 & 0.256 \\
N=71,863 & (0.003) &  & (0.003) & (0.003) & (0.003) & (0.003) & (0.003) & (0.003) \\
\# vars &  &  & 76 & 234 & 234 & 234 & 234 & 234 \\
\# vars selected &  &  & 76 & 14 & 51 & 105 & 173 & 44 \\[.1cm]
Age 25-40 & 0.219 &  & 0.238 & 0.181 & 0.233 & 0.244 & 0.240 & 0.257 \\
N=71,871 & (0.003) &  & (0.003) & (0.003) & (0.003) & (0.003) & (0.003) & (0.003) \\
\# vars &  &  & 76 & 233 & 233 & 233 & 233 & 233 \\
\# vars selected &  &  & 76 & 20 & 60 & 112 & 180 & 43 \\[.1cm]
Age 25-45 & 0.219 &  & 0.237 & 0.213 & 0.243 & 0.240 & 0.235 & 0.245 \\
N=71,873 & (0.003) &  & (0.003) & (0.003) & (0.003) & (0.003) & (0.003) & (0.003) \\
\# vars &  &  & 76 & 234 & 234 & 234 & 234 & 234 \\
\# vars selected & &  & 76 & 17 & 63 & 121 & 181 & 45 \\
    \bottomrule
    \bottomrule
    \end{tabular}%
    \end{adjustbox}
    \smallskip
  \fnote{\textbf{Notes:} The table reports the slope coefficient from a regression of son’s income on father’s lifetime income. The measure for son’s income is lifetime income in column (1) or the predicted lifetime income from a first-step estimation of equation (\ref{eqtwostepFE}) in columns (2)-(7), using our preferred predictors in column (2) or selecting predictors by lasso in columns (3)-(7). In column (7) we include the parental income x child age interactions as non-penalized regressors in the first step. See text for a description of each estimator. Standard errors in parentheses.}
  \label{tab:AppendixML100p}%
\end{table}%

Table \ref{tab:AppendixML100p} compares estimates of the IGE using our preferred first-step estimator (``parental quadratic'', column 2) with those based on lasso to select the first-step predictors. To show how the performance of the latter varies with the number of selected predictors, we vary the lasso tuning parameter $\lambda$ (columns 3-6). We include a broad range of candidate predictors: alongside our standard variables (e.g. child education, parental income) we also consider family size, birth order, an indicator for second-generation immigrants, cognitive and non-cognitive skill scores (as also used in Table 2), and include all two-way interactions of these variables with child age and with age squared, resulting in 233 candidate variables. We also account for individual fixed effects in all specifications.\footnote{Since available implementations of lasso and elastic net in Stata or R struggle with handling many (non-penalized) fixed effects, we first residualized the outcome and all potential predictors against these fixed effects and then used Stata's lasso and elasticnet commands with the residualized variables. This approach is numerically equivalent (Frisch-Waugh-Lovell theorem), as we also verified in our data.} As the skill scores contain missings, the sample for our ML analysis is smaller than the main intergenerational sample used in the rest of the paper, resulting in a slightly smaller benchmark estimate of the IGE (column 1).

When observing child income only at age 25-27, the lasso-based estimator tends to perform worse than our preferred parametric estimator, especially when selecting only a limited set of predictors (corresponding to a larger $\lambda$). For instance, the IGE estimate is only 0.117 when selecting 48 predictors (column 4), rising to 0.173 with 105 non-zero predictors (column 5). This is substantially below the benchmark estimate (0.219, column 1), and also lower than the estimate from our parametric approach with 76 first-stage predictors (0.203, column 2). That is, despite selecting more predictors for the estimation of income profiles than our preferred specification, the lasso estimator results in a greater downward bias in the IGE. All lifecyle estimators perform better when child income is observed over wider age ranges, and the gap between our preferred parametric and the lasso-based estimators becomes less pronounced. Still, the lasso tends to perform worse than our parametric procedure. 

The reason why the lasso-based approach performs worse is related to both the variable selection and regularization steps that it entails. First, lasso does not reliably select the interactions between child age and parental income that, from Figure \ref{fig:fig1}, we understand to be a crucial source of bias in the second-step estimation of the IGE. When including these interactions as non-penalized regressors, the lasso performs similarly as our parametric estimator (``not penalized'' lasso, column 7). The lasso also performs well when selecting a very small value for $\lambda$ (column 6), as the resulting longer list of predictors then also includes the crucial interactions between child age and parental income. The lasso does not reliably select these predictors as it maximizes the predictive accuracy in the (first-step) prediction of lifetime incomes, whereas our objective is to reduce bias in the (second-step) estimation of the IGE. Although the two objectives are related, the key source of bias in the IGE –– variation in income growth rates by parental income (see Figure \ref{fig:fig1}) -- may not be a particularly strong predictor of lifetime incomes, as also noted in Section \ref{sec:newcorrection}.

\begin{table}[th!]
\centering
  \caption{ML Estimation of Lifecycle Profiles (postselection OLS)}
     \begin{adjustbox}{width=\textwidth}
         \begin{tabular}{lcccccccc}
    \toprule
    \toprule
          & \multicolumn{1}{c}{Direct} &       & \multicolumn{6}{c}{Lifecycle estimator} \\
\cmidrule{2-2}\cmidrule{4-9}          & Lifetime &       & Parental & Lasso (OLS) & Lasso (OLS) & Lasso (OLS) & Lasso (OLS) & Lasso (OLS) \\
          &       &       &  Quadratic & $\lambda=0.01$ & $\lambda=0.001$ & $\lambda=0.0001$ & $\lambda=0.00001$ & $\lambda=0.01$ \\
          &       &       &    &  &  &  &  & not pen. \\        
    Son's Age & (1)  & & (2)  & (3)   & (4)   & (5)   & (6)   & (7) \\
    \midrule
Age 25-27 & 0.219 &  & 0.203 & 0.058 & 0.152 & 0.202 & 0.224 & 0.201 \\
N=71,794 & (0.003) &  & (0.004) & (0.004) & (0.004) & (0.004) & (0.004) & (0.004) \\
\# vars &  &  & 76 & 233 & 233 & 233 & 233 & 233 \\
\# vars selected &  &  & 76 & 16 & 48 & 105 & 174 & 42 \\[.1cm]
Age 25-30 & 0.219 &  & 0.241 & 0.133 & 0.217 & 0.253 & 0.255 & 0.271 \\
N=71,846 & (0.003) &  & (0.004) & (0.003) & (0.004) & (0.004) & (0.004) & (0.004) \\
\# vars &  &  & 76 & 234 & 234 & 234 & 234 & 234 \\
\# vars selected &  &  & 76 & 15 & 52 & 115 & 170 & 45 \\[.1cm]
Age 25-35 & 0.219 &  & 0.229 & 0.169 & 0.226 & 0.235 & 0.231 & 0.255 \\
N=71,863 & (0.003) &  & (0.003) & (0.003) & (0.003) & (0.003) & (0.003) & (0.003) \\
\# vars &  &  & 76 & 234 & 234 & 234 & 234 & 234 \\
\# vars selected &  &  & 76 & 14 & 51 & 105 & 173 & 44 \\[.1cm]
Age 25-40 & 0.219 &  & 0.238 & 0.207 & 0.245 & 0.242 & 0.238 & 0.256 \\
N=71,871 & (0.003) &  & (0.003) & (0.003) & (0.003) & (0.003) & (0.003) & (0.003) \\
\# vars &  &  & 76 & 233 & 233 & 233 & 233 & 233 \\
\# vars selected &  &  & 76 & 20 & 60 & 112 & 180 & 43 \\[.1cm]
Age 25-45 & 0.219 &  & 0.237 & 0.233 & 0.247 & 0.233 & 0.236 & 0.245 \\
N=71,873 & (0.003) &  & (0.003) & (0.004) & (0.003) & (0.003) & (0.003) & (0.003) \\
\# vars &  &  & 76 & 234 & 234 & 234 & 234 & 234 \\
\# vars selected & &  & 76 & 17 & 63 & 121 & 181 & 45 \\
    \bottomrule
    \bottomrule
    \end{tabular}%
    \end{adjustbox}
    \smallskip
  \fnote{\textbf{Notes:} The table reports the slope coefficient from a regression of son’s income on father’s lifetime income. The measure for son’s income is lifetime income in column (1) or the predicted lifetime income from a first-step estimation of equation (\ref{eqtwostepFE}) in columns (2)-(7). In columns (3)-(7) we use the postselection OLS rather than penalized lasso coefficients. In column (7) we include the parental income x child age interactions as non-penalized regressors in the first step. See text for a description of each estimator. Standard errors in parentheses.}
  \label{tab:AppendixML100pOLS}%
\end{table}%

Another source of bias in the lasso-based approach is the regularization (or shrinkage) of coefficients. While improving out-of-sample predictive accuracy, regularization in the first step may induce bias in the second step (\citealt{chernozhukov2022locally}). To probe this issue, Table \ref{tab:AppendixML100pOLS} reports the resulting estimates of IGE when using the ``postselection'' rather than penalized lasso coefficients in the first-step prediction of income profiles, which are calculated by taking the variables selected by lasso and refitting the model by OLS. While this improves the second-step estimates of the IGE, they are still substantially biased when child income is only observed at young ages, unless a sufficiently large set of predictors is selected. For example, when observing child income between age 25 and 27, using postselection lasso in the first step with 48 predictors yields an estimate of the IGE of only 0.152 compared to the benchmark estimate of 0.219.

\begin{table}[ht!]
\centering
  \caption{ML Estimation of Lifecycle Profiles (2\% sample)}
     \begin{adjustbox}{width=\textwidth}
         \begin{tabular}{lcccccccc}
    \toprule
    \toprule
          & \multicolumn{1}{c}{Direct} &       & \multicolumn{6}{c}{Lifecycle estimator} \\
\cmidrule{2-2}\cmidrule{4-9}          & Lifetime &       & Parental & Lasso & Lasso & Elastic net & Elastic net & Lasso \\
          &       &       &  Quadratic &  & postselection  &  & postselection & not pen. \\        
    Son's Age & (1)  & & (2)  & (3)   & (4)   & (5)   & (6)   & (7) \\
    \midrule
Age 25-27 & 0.212 &  & 0.188 & 0.131 & 0.156 & 0.131 & 0.156 & 0.198 \\
N=5,686 & (0.013) &  & (0.014) & (0.014) & (0.014) & (0.014) & (0.014) & (0.014) \\
$\lambda$ &  &  &  & 0.00032 & 0.00032 & 0.00032 & 0.00032 & 0.00032 \\
$\alpha$ &  &  &  & 1 & 1 & 1 & 1 & 1 \\
\# vars &  &  & 76 & 234 & 234 & 234 & 234 & 234 \\
\# vars selected &   &  & 76 & 97 & 97 & 97 & 97 & 108 \\[.1cm]
Age 25-30 & 0.212 &  & 0.224 & 0.224 & 0.226 & 0.223 & 0.225 & 0.257 \\
N=5,688 & (0.013) &  & (0.012) & (0.013) & (0.013) & (0.013) & (0.013) & (0.013) \\
$\lambda$ &  &  &  & 0.00010 & 0.00010 & 0.00012 & 0.00012 & 0.00054 \\
$\alpha$ &  &  &  & 1 & 1 & 0.75 & 0.75 & 1 \\
\# vars &  &  & 76 & 234 & 234 & 234 & 234 & 234 \\
\# vars selected &    &  & 76 & 122 & 122 & 126 & 126 & 96 \\[.1cm]
Age 25-35 & 0.212 &  & 0.214 & 0.217 & 0.232 & 0.217 & 0.232 & 0.250 \\
N=5,694 & (0.013) &  & (0.011) & (0.012) & (0.012) & (0.012) & (0.012) & (0.012) \\
$\lambda$ &  &  &  & 0.00054 & 0.00054 & 0.00054 & 0.00054 & 0.00558 \\
$\alpha$ &  &  &  & 1 & 1 & 1 & 1 & 1 \\
\# vars &  &  & 76 & 234 & 234 & 234 & 234 & 234 \\
\# vars selected &   &  & 76 & 93 & 93 & 93 & 93 & 48 \\[.1cm]
Age 25-40 & 0.212 &  & 0.243 & 0.257 & 0.261 & 0.257 & 0.262 & 0.270 \\
N=5,694 & (0.013) &  & (0.012) & (0.012) & (0.012) & (0.012) & (0.012) & (0.012) \\
$\lambda$ &  &  &  & 0.00068 & 0.00068 & 0.00129 & 0.00129 & 0.00210 \\
$\alpha$ &  &  &  & 1 & 1 & 0.5 & 0.5 & 1 \\
\# vars &  &  & 76 & 234 & 234 & 234 & 234 & 234 \\
\# vars selected &   &  & 76 & 85 & 85 & 87 & 87 & 66 \\[.1cm]
Age 25-45 & 0.212 &  & 0.242 & 0.253 & 0.254 & 0.253 & 0.254 & 0.256 \\
N=5,694 & (0.013) &  & (0.012) & (0.012) & (0.012) & (0.012) & (0.012) & (0.012) \\
$\lambda$ &  &  &  & 0.00052 & 0.00052 & 0.00100 & 0.00100 & 0.00270 \\
$\alpha$ &  &  &  & 1 & 1 & 0.5 & 0.5 & 1 \\
\# vars &  &  & 76 & 234 & 234 & 234 & 234 & 234 \\
\# vars selected &  &  & 76 & 87 & 87 & 88 & 88 & 58 \\
    \bottomrule
    \bottomrule
    \end{tabular}%
    \end{adjustbox}
    \smallskip
  \fnote{\textbf{Notes:} The table reports the slope coefficient from a regression of son’s income on father’s lifetime income. The measure for son’s income is lifetime income in column (1) or the predicted lifetime income from a first-step estimation of equation (\ref{eqtwostepFE}) in columns (3)-(7), using the indicated method to select the predictors. In column (7) we include the parental income x child age interactions as non-penalized regressors in the first step. See text for a description of each estimator. Standard errors in parentheses.}
  \label{tab:AppendixML2p}%
\end{table}%

Still, ML methods may perform well in large samples such as ours, if a low tuning parameter $\lambda$ and therefore sufficiently many predictor variables are chosen (see column 6 of Tables \ref{tab:AppendixML100p} and Table \ref{tab:AppendixML100pOLS}). In contrast, ML estimation of the first step will yield worse results in smaller samples. To illustrate this, Table \ref{tab:AppendixML2p} summarizes the performance of ML-based lifecycle estimators in a 2\% subsample of our main intergenerational sample. We consider both lasso and elastic net, using either postselection or penalized coefficients for predicting lifetime incomes in the first step. We use cross-validation to select the optimal tuning parameters $\lambda$ (lasso) and $\alpha$ (elastic net), selecting between $\alpha=0.5$, $0.75$, or $1$ (where $\alpha=1$ corresponds to the lasso). As even our 2\% subsample contains several thousand individuals, the set of selected predictors remains large (between 85 and 126 predictors, depending on specification). Still, the ML-based lifecycle estimators tend to perform worse than our preferred parametric specification. For example, when observing child income between age 25 and 27, using the penalized lasso coefficients to predict lifetime incomes yields a second-step estimate of the IGE of 0.131, substantially below the benchmark (0.212, column 1) or our preferred lifecycle estimator (0.188, column 2). As before, including the interactions between child age and parental income as non-penalized regressors improves the performance of the lasso-based estimator (column 7). 

We therefore conclude that plugging in standard machine learning methods in the first step is unlikely to lead to better IGE estimates than our parametric approach. Of course, this does not rule out the possibility that more tailored applications of ML methods may prove useful. Plug-in ML methods perform poorly because they aim to maximize predictive accuracy in the first-step estimation of lifecycle income profiles, rather than to minimize bias in the second-step estimation of the IGE (i.e. they target the wrong objective function); regularization and model selection bias from the first step then generate bias in the second step. More sophisticated ML implementations that explicitly minimize bias in the second step may perform better than plug-in ML methods, but would also be harder for practitioners to implement. One promising strategy considered by \cite{Puerta2024RobustIGE} is to construct a debiased machine learning estimator for the IGE based on \textit{orthogonal} moments (\citealt{chernozhukov2022locally}). 

\end{document}